\documentclass{article}

\usepackage{PRIMEarxiv}

\usepackage[utf8]{inputenc} 
\usepackage[T1]{fontenc}    
\usepackage{hyperref}       
\usepackage{url}   
\usepackage{acro}
\DeclareAcronym{ecc}{
short = ECC,
long = error correcting code
}
\DeclareAcronym{ber}{
short = BER,
long  = bit error rate
}
\DeclareAcronym{asic}{
short        = ASIC,
long         = application-specific integrated-circuit
}
\DeclareAcronym{fpga}{
short = FPGA,
long  = field-programmable gate array
}
\DeclareAcronym{gpu}{
short = GPU,
long  = graphics processing unit
}
\DeclareAcronym{ldpc}{
short = LDPC,
long = low-density parity-check
}
\DeclareAcronym{nbldpc}{
short = NB-LDPC,
long = non-binary low-density parity-check
}
\DeclareAcronym{dvbs2}{
short = DVB-S2,
long = ETSI digital video broadcasting \nth{2} generation
}
\DeclareAcronym{cn}{
short = CN,
long = check node
}
\DeclareAcronym{vn}{
short = VN,
long = variable node
}
\DeclareAcronym{spa}{
short = SPA,
long = sum-product algorithm
}
\DeclareAcronym{lspa}{
short = LSPA,
long = log sum-product algorithm
}
\DeclareAcronym{emsa}{
short = EMSA,
long = extended min-sum algorithm
}
\DeclareAcronym{mma}{
short = MMA,
long = min-max algorithm
}
\DeclareAcronym{pcm}{
short = PCM,
long = parity-check matrix
}
\DeclareAcronym{bn}{
short = BN,
long = bit node
}
\DeclareAcronym{qcldpc}{
short = QC-LDPC,
long = quasi-cyclic low-density parity-check
}
\DeclareAcronym{gf}{
short = GF,
long = Galois field
}
\DeclareAcronym{mlgd}{
short = MLDG,
long =  majority-logic decoding
}
\DeclareAcronym{ihrb}{
short = IHRB,
long =  iterative hard reliability-based
}
\DeclareAcronym{isrb}{
short = ISRB,
long =  iterative soft reliability-based
}
\DeclareAcronym{gbfda}{
short = GBFDA,
long =  generalized bit-flipping decoding algorithm
}
\DeclareAcronym{amsa}{
short = AMSA,
long =  Adaptive Multiset Stochastic Algorithm
}
\DeclareAcronym{wbrb}{
short = WBRB,
long =  weighted bit reliability-based
}
\DeclareAcronym{fbrb}{
short = FBRB,
long =  full bit reliability-based
}
\DeclareAcronym{mimo}{
short = MIMO,
long =  multiple-input multiple-output
}
\DeclareAcronym{qam}{
short = QAM,
long =  quadrature amplitude modulation
}
\DeclareAcronym{snr}{
short = SNR,
long =  signal-to-noise ratio
}
\DeclareAcronym{adbp}{
short = ADBP,
long =  analog digital belief propagation
}
\DeclareAcronym{srb}{
short = SRB,
long =  symbol reliability based
}
\DeclareAcronym{gps}{
short = GPS,
long =  global positioning system
}
\DeclareAcronym{vlsi}{
short = VLSI,
long =  very large scale integration
}
\DeclareAcronym{fht}{
short = FHT,
long =  fast Hadamard transform
}
\DeclareAcronym{fft}{
short = FFT,
long =  fast Fourier transform
}
\DeclareAcronym{hls}{
short = HLS,
long =  high-level synthesis
}
\DeclareAcronym{rtl}{
short = RTL,
long =  register transfer level
}

\DeclareAcronym{qos}{
short = QoS,
long =  quality of service
}

\DeclareAcronym{bp}{
short = BP,
long =  belief propagation
}

\DeclareAcronym{llr}{
short = LLR,
long =  log-likelihood ratio
}

\DeclareAcronym{csr}{
short = CSR,
long =  compressed sparse row
}

\DeclareAcronym{csc}{
short = CSC,
long =  compressed sparse column
}

\DeclareAcronym{lut}{
short = LUT,
long =  look-up table
}

\DeclareAcronym{cpu}{
short = CPU,
long  = central processing unit
}

\DeclareAcronym{sm}{
short = SM,
long  = streaming multiprocessor
}

\DeclareAcronym{ram}{
short = RAM,
long  = random access memory
}

\DeclareAcronym{ccsds}{
short = CCSDS,
long  = Consultative Committee for Space Data Systems
}

\DeclareAcronym{bch}{
short = BCH,
long  = Bose–Chaudhuri–Hocquenghem
}

\DeclareAcronym{fwht}{
short = FWHT,
long  = fast Walsh-Hadamard transform
}

\DeclareAcronym{tdp}{
short = TDP,
long  = thermal design power
}

\DeclareAcronym{cuda}{
short = CUDA,
long  = compute unified device architecture
}

\DeclareAcronym{pim}{
short = PiM,
long  = processing-in-memory
}

\DeclareAcronym{pnm}{
short = PnM,
long  = processing-near-memory
}

\DeclareAcronym{pum}{
short = PuM,
long  = processing-using-memory
}

\DeclareAcronym{dpu}{
short = DPU,
long  = DRAM processing unit
}

\DeclareAcronym{ms}{
short = MS,
long = min-sum
}

\DeclareAcronym{wram}{
short = WRAM,
long = working SRAM
}

\DeclareAcronym{mram}{
short = MRAM,
long = main DRAM
}

\DeclareAcronym{sram}{
short = SRAM,
long =static random-access memory
}

\DeclareAcronym{dram}{
short = DRAM,
long =dynamic random-access memory
}

\DeclareAcronym{mm}{
short = MM,
long =min-max
}

\DeclareAcronym{alu}{
short = ALU,
long = arithmetic logic unit
}

\DeclareAcronym{bpsk}{
short = BPSK,
long = binary phase-shift keying
}

\DeclareAcronym{awgn}{
short = AWGN,
long = additive white Gaussian noise
}

\DeclareAcronym{bf}{
short = BF,
long = bit-flipping
}

\DeclareAcronym{cnp}{
short = CNP,
long = check node processing
}

\DeclareAcronym{vnp}{
short = VNP,
long = variable node processing
}

\DeclareAcronym{app}{
short = APP,
long = \textit{a posteriori} probability
}

\DeclareAcronym{isa}{
short = ISA,
long = instruction set architecture
}

\DeclareAcronym{risc}{
short = RISC,
long = reduced instruction set computer
}

\DeclareAcronym{iram}{
short = IRAM,
long = instruction DRAM
}

\DeclareAcronym{fp}{
short = FP,
long = floating-point
}

\DeclareAcronym{dma}{
short = DMA,
long = direct memory access
}

\DeclareAcronym{dimm}{
short = DIMM,
long = dual in-line memory module
}

\DeclareAcronym{ddr4}{
short = DDR4,
long = double data rate 4
}

\DeclareAcronym{spmd}{
short = SPMD,
long = {single program, multiple data}
}

\DeclareAcronym{sdk}{
short = SDK,
long = software development kit
}

\DeclareAcronym{simd}{
short = SIMD,
long = {single instruction, multiple data}
}

\DeclareAcronym{simt}{
short = SIMT,
long = {single instruction, multiple thread}
}

\DeclareAcronym{raw}{
short = RAW,
long = {read-after-write}
}

\DeclareAcronym{ran}{
short = RAN,
long = {radio access network}
}

\usepackage{booktabs}       
\usepackage{amsmath,amssymb,amsfonts}
\usepackage{nicefrac}       
\usepackage{microtype}      
\usepackage{lipsum}
\usepackage{fancyhdr}       
\usepackage{graphicx}       
\graphicspath{{media/}}     
\usepackage{algorithmic}
\usepackage{algorithm}
\usepackage{comment}
\usepackage{multirow}
\usepackage{xcolor}
\usepackage{soul}
\usepackage{bm}
\usepackage{tcolorbox}
\title{In-Memory Non-Binary LDPC Decoding
}

\author{
  Oscar Ferraz, Vitor Silva, and Gabriel Falcao\\
  University of Coimbra\\
Instituto de Telecomunicações\\ and the CUDA Research Center in Coimbra\\
Department of Electrical and Computer Engineering\\ Coimbra 3030-290\\
Portugal\\
  \texttt{\{oscar.ferraz, vitor, gff\}@co.it.pt} \\
}

\begin{document}
\maketitle

\begin{abstract}
Low-density parity-check (LDPC) codes are an important feature of several communication and storage applications, offering a flexible and effective method for error correction. These codes are computationally complex and require the exploitation of parallel processing to meet real-time constraints. As advancements in arithmetic and logic unit technology allowed for higher performance of computing systems, memory technology has not kept the same pace of development, creating a data movement bottleneck and affecting parallel processing systems more dramatically. To alleviate the severity of this bottleneck, several solutions have been proposed, namely the processing in-memory (PiM) paradigm that involves the design of compute units to where (or near) the data is stored, utilizing thousands of low-complexity processing units to perform out bit-wise and simple arithmetic operations. This paper presents a novel efficient solution for near-memory non-binary LDPC decoders in the UPMEM system, for the best of our knowledge the first real hardware PiM-based non-binary LDPC decoder that is benchmarked against low-power GPU parallel solutions highly optimized for throughput performance. PiM-based non-binary LDPC decoders can achieve $76$ Mbit/s of decoding throughput, which is even competitive when compared against implementations running in edge GPUs.
\end{abstract}

\keywords{LDPC Codes \and Non-binary LDPC Decoding \and Parallel Signal Processing \and Processing in-Memory \and UPMEM.}

\section{Introduction}
\label{sec:introduction}
\Ac{ldpc} codes are important components in the communications~\cite{CCSDS:2015, Kang:2020, Lam:2022, Xie:2021}, and data storage~\cite{Mondal:2021, Zhang:2023} fields. The strong error-correcting capability of \ac{ldpc} codes allows them to approach the Shannon limit~\cite{Davey:1998}, and, more recently, they have been adopted in the $5$G New Radio technical specification~\cite{Holma_2020, Le:2021}. Nevertheless, developing efficient \ac{ldpc} decoders presents significant challenges due to their demanding computational requirements and memory access patterns. Over the years, researchers have proposed parallel approaches on different hardware platforms, including \acp{cpu}~\cite{Gal:2017, le2020high}, \acp{gpu}~\cite{Wang:2012, ferraz:2021:asilomar}, \acp{fpga}~\cite{Liu:2021, Nadal:2021}
, and \acp{asic}~\cite{Ferraz:2021, Nguyen:2017} to improve throughput performance and meet latency constraints. However, the partial-parallel nature of these decoders still faces challenges in the computations of the partial results, which, in turn, increases data transfers between computational units and memory.

\Ac{ldpc} codes were proposed in $1962$ by Robert Gallager~\cite{Gallager:1962} but were considered impractical due to their computational complexity. In $1998$, Davey and MacKay proposed the \ac{nbldpc} codes as an extension to the binary \ac{ldpc} codes that allowed to achieve better error-correction performance for moderate code lengths at the cost of higher computational complexity~\cite{Davey:1998, Feng:2018}. In the last two decades, the development of more efficient computational systems allowed the implementation of \ac{ldpc} decoders in error correction applications using parallel computing architectures such as \acp{cpu}, \acp{gpu}, \acp{fpga}, and \acp{asic}. These platforms allowed decoders to increase their throughput performance. However, when choosing a platform to compute such a highly complex subclass of \ac{ldpc} codes, a trade-off between throughput, latency, memory accesses, energy consumption, error-correction capability, flexibility, and cost must be made~\cite{Ferraz:2021, Shao:2019}.


Over the past few decades, the development of computing units has adhered to Moore's law, where the transistor count doubles approximately every $18$ months for the same area. However, challenges in \ac{dram} scaling (increasing density and performance while maintaining reliability, latency, and energy consumption) have caused memory technology to not match the same pace of development as computational units, thus leading to a \textit{"data movement bottleneck"}~\cite{Ghiasi:2023, Mutlu:2022}. This bottleneck is present in most modern computing systems. It happens when a significant part of the execution time is spent on data transfers between memory and processing units. 


These obstacles in improving \ac{dram} technology have motivated the industry and researchers to rethink and redesign memory systems~\cite{Hosseini:2022}, such as $3$D stacked~\cite{Sutradhar:2024} and non-volatile memories~\cite{Zhang:2023:survey}. Several new approaches propose a paradigm shift from a processor-centric design to a \ac{pim} design~\cite{Li:2023} where the memory units are designed to incorporate computational capability, allowing to alleviate the \textit{"data movement bottleneck"} by computing data where (or near) it resides~\cite{ Mutlu:2022, Li:2023, Zhang:2023:survey}.

 Currently, \ac{pim} systems  can be divided into two main areas:

\begin{itemize}
  \item \textbf{\protect\Ac{pnm}} incorporates computing units near the memory, such as $3$D stacked memory, which integrates several memory layers with one (or more) logic layer(s)~\cite{gomez:2021}. \Ac{pnm} supports a large range of operations. However, the main downside is that these systems suffer from area and thermal constraints, have limited capacity, and have a high cost~\cite{gomez:2021}.

  \item \textbf{\protect\Ac{pum}} takes advantage of the analog circuit properties of memory cells (\ac{sram}~\cite{Fujiki:2019}, \ac{dram}~\cite{seshadri:2016}, and non-volatile memories~\cite{angizi:2018}) to perform simple bitwise operations (AND and OR operations)~\cite{Seshadri:2017}. Although very efficient in terms of performance, the drawback is that this paradigm requires a major reorganization of memory circuitry to enable more complex operations to be performed~\cite{Seshadri:2017}.
\end{itemize}

Due to the challenges mentioned above, the industry is still designing computing systems with \ac{pim} technology. The UPMEM \ac{pim} architecture is the first commercial realization of a \ac{pnm} system implemented in hardware~\cite{UPMEM:2018, gomez:2021}. This system is a massive multicore-\ac{cpu} with private memory that takes advantage of the more mature fabrication and design of $2$D \ac{dram} to incorporate computational units in the same chip~\cite{UPMEM:2018, gomez:2021}, allowing the implementation of several independent \ac{dpu} cores with deep pipelines ($14$ stages) and fine-grained multithreading (up to $24$ threads)~\cite{UPMEM:2018}.


To address the performance bottlenecks associated with data transfers in LDPC decoding, this paper proposes the implementation of \ac{ldpc} decoders in \ac{pnm} architectures. Integrating \ac{ldpc} decoding into \ac{pim} can significantly alleviate the data movement bottleneck between memory and compute units, thereby enhancing overall performance. In the UPMEM system, we leverage the advantages of the memory hierarchy (\ac{mram}, \ac{wram}) to minimize data transfers, particularly in scenarios where irregular memory access patterns pose significant challenges. We also take advantage of multithreading and $8$-bit multipliers supported by the simple \ac{alu} across several \acp{dpu} to maximize throughput performance. By exploiting \ac{pnm} architectures, this paper aims to demonstrate how leveraging \ac{pim} technology can effectively mitigate the performance limitations encountered in typical parallel \ac{ldpc} decoding implementations, ultimately enabling substantial improvements in real-time processing capabilities.

While \ac{nbldpc} decoders have been extensively explored on \acp{cpu}, \acp{gpu}, and \acp{fpga}~\cite{Ferraz:2021}, their implementation on \ac{pim} architectures remains largely unexplored. Therefore, \textbf{this paper proposes the following contributions:}

\begin{itemize}
 \item \textbf{The first \ac{pim}-based \ac{nbldpc}} \Ac{fft}-\Ac{spa} and \Ac{mm} \textbf{decoder implementations}, to the best of our knowledge.
 
    \item An \textbf{efficient mapping of the \ac{nbldpc} decoders} to the UPMEM architecture, exploiting the system's memory hierarchy (\ac{mram}, \ac{wram}), multithreading, multicodeword decoding, and $8$-bit multipliers, in order to avoid the UPMEM's processing limitations for \ac{fp} operations.

    \item \textbf{Performance comparison} against \textbf{low-power} parallel architectures, such as embedded \textbf{\protect\acp{gpu}}, using real codes recommended by the \ac{ccsds}-$231$~\cite{ccsds231}.

\end{itemize}

\section{Belief Propagation}

Belief propagation is an iterative message-passing algorithm used in graphical models to infer properties of a small data set from the observation of a larger data set. This algorithm is successfully applied in information theory, artificial  intelligence, computer vision, and decision support systems, with the principal applications of belief propagation being \ac{ldpc} codes~\cite{Gallager:1962}, turbo codes~\cite{Shen:2023}, and Bayesian  networks~\cite{Pearl:2022}.

\begin{figure}[!t]
  \centering
  \includegraphics[width=.7\columnwidth]{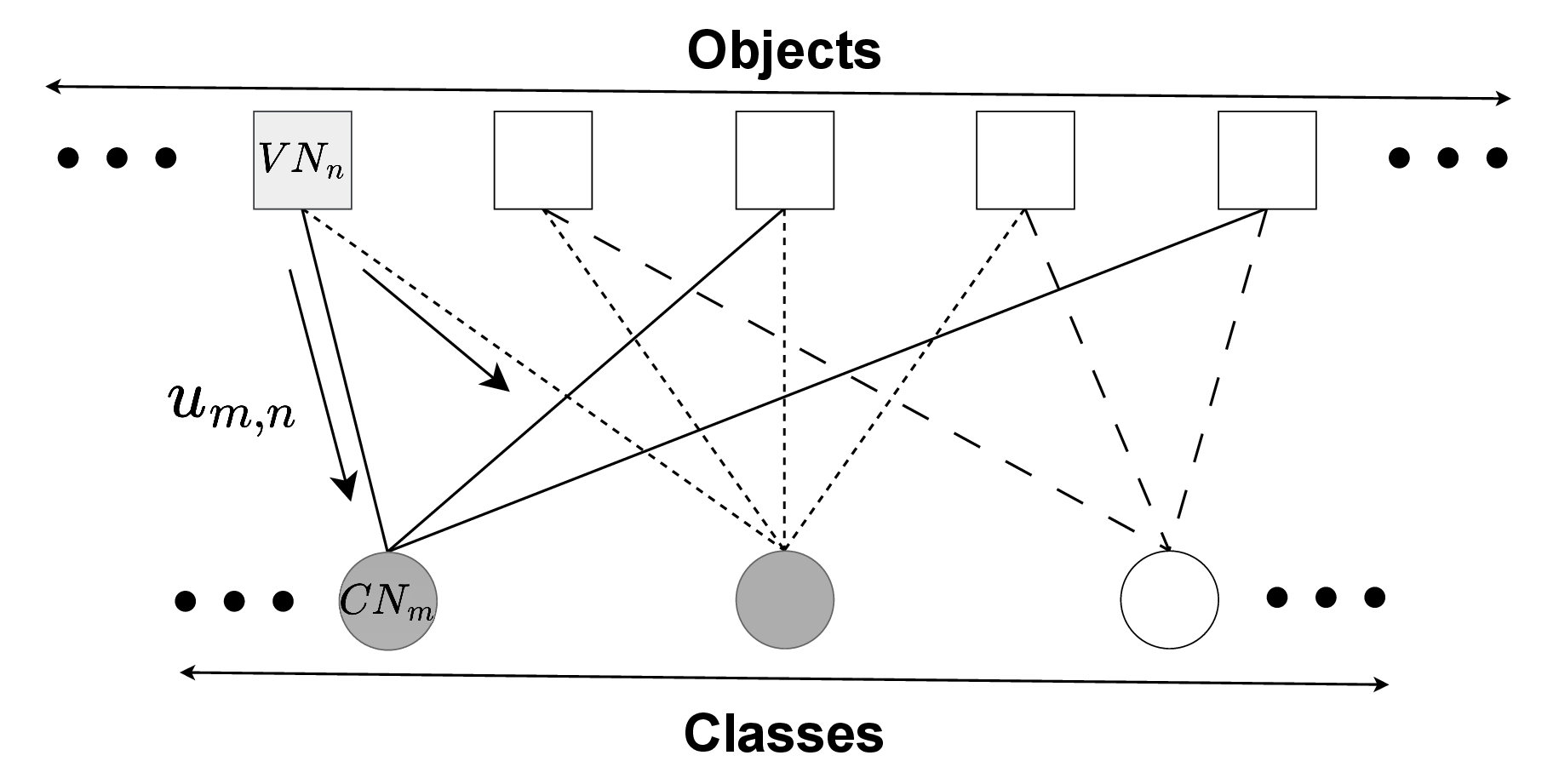}
  \caption{A bipartite graph depicting relationships between objects and their classes. In the graph, belief propagation refines the probability of an object belonging to a class depending on the probability of other objects belonging to the same class. The probabilities between nodes $n$ and $m$ are exchanged in the form of messages $u_{m, n}$.}
  \label{fig:belief_propagation}
\end{figure}

Graphs are structures that represent relationships between objects (also called nodes). The belief propagation algorithm calculates the probability distribution of a subset of nodes. The information gathered from calculated distributions can be used to infer the properties of objects  from other subsets that maintain relationships between them. For instance, a belief network is used in a medical diagnosis support system to establish  relationships between symptoms and diseases~\cite{Nassim:2014} or in image classification~\cite{cardoso:2024}, as depicted in Fig.~\ref{fig:belief_propagation}.

In the case of \ac{ldpc} codes, a bipartite graph describes the mathematical relationships between several nodes, where probabilities are exchanged between them.

\subsection{Non-binary LDPC Codes}

Noise is an inherent and undesirable part of most communication systems. In digital data communications, data is transmitted from the source to a receiver through a channel. However, the channel is susceptible to noise that corrupts and changes bits during transmission, requiring the deployment of \acp{ecc} (for instance, \ac{ldpc} codes) in the receiver to detect and correct the erroneous bits. This provides resilience to noise in data.



\begin{figure}[!t]
  \centering
  \includegraphics[width=0.9\columnwidth]{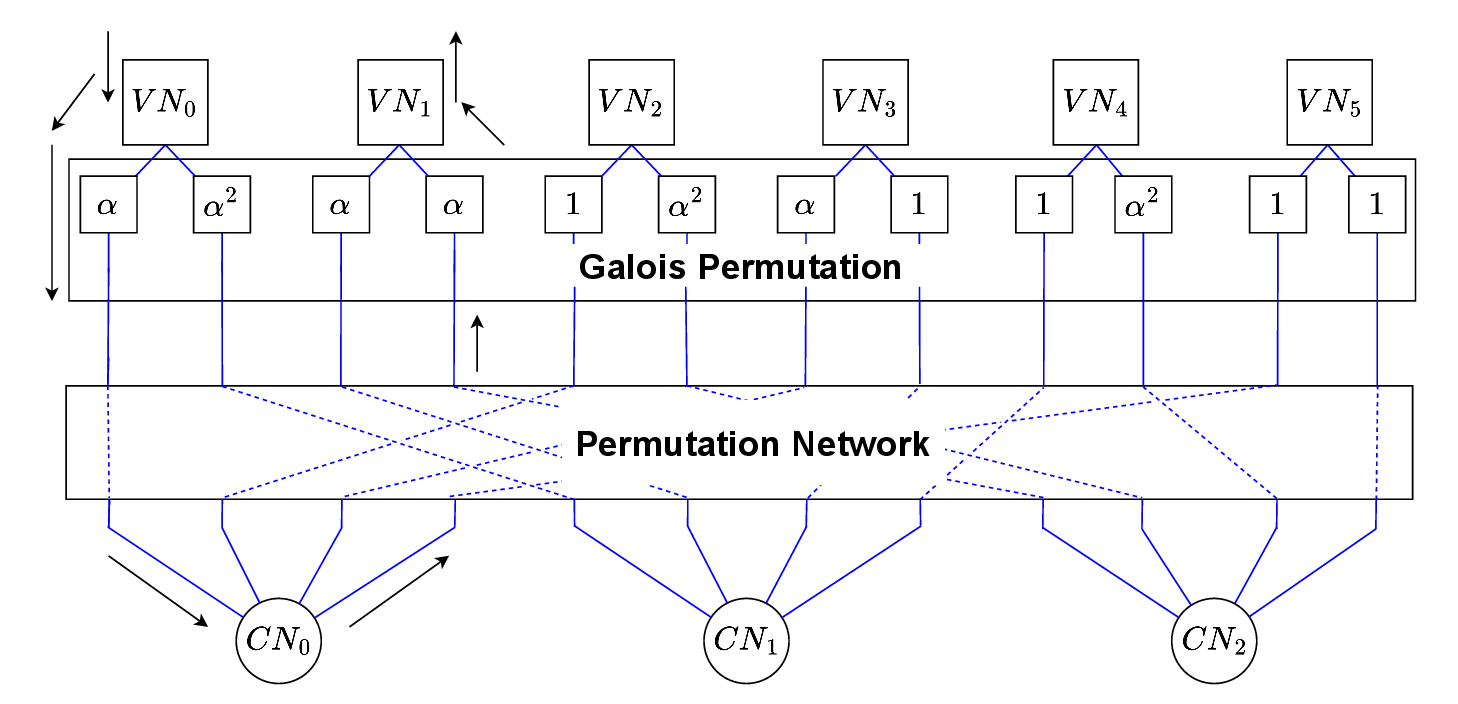}
  \caption{The \ac{nbldpc} Tanner graph representation of matrix~(\ref{eq:NBPCM}). The number of \acp{vn} ($N$) corresponds to the number of columns in the $\mathbf{H}$ matrix, while the number of \acp{cn} ($M$) matches the number of rows. The permutation network is defined by the non-zero entries in the $\mathbf{H}$ matrix. The probability of a symbol being right or wrong is transmitted through the connected nodes and processed by \acp{vn} and \acp{cn}. 
  }
  \label{fig:tanner_graph}
\end{figure}

An \ac{nbldpc} code is a linear block code defined by the sparse \ac{pcm} $\mathbf{H}$ of size $M$ by $N$, in $GF(g)$  or a Tanner graph representation, as illustrated in Fig.~\ref{fig:belief_propagation} and~\ref{fig:tanner_graph}~\cite{carrasco:2008}. $M$ represents the number of rows of $\mathbf{H}$ or the number of \acp{cn} in the Tanner graph, while $N$ indicates the number of columns or the number of \acp{vn}. For instance, in~(\ref*{eq:NBPCM}), $M=3$ rows or three \acp{cn} and $N=6$ columns  or six \acp{vn}. The connection between \acp{cn} and \acp{vn} is established  by the non-zero elements of $\mathbf{H}$, in~(\ref{eq:NBPCM}), the $CN_0$ is connected to $VN_0$, and the $CN_0$ is not connected to $VN_1$. An \ac{ldpc} code is considered regular when each row contains the same number of non-zero elements (\ac{cn} degree), and each column also contains an equal number of non-null elements (\ac{vn} degree). Conversely, if the \ac{cn} degree ($d_c$) and the \ac{vn} degree ($d_v$) varies, the code is considered irregular. Irregular codes offer improved coding performance but require higher computational complexity.

Barnault and Declercq proposed \ac{nbldpc} that can be decoded using \acp{gf} up to $GF(2^q),$ $q\in \mathbb{Z}^+,$ $q \leq 8$~\cite{Barnault:2003}. A \ac{gf} is a mathematical field with  finite elements, represented as $GF(g)$, where $g$ indicates the field's cardinality~\cite{Barnault:2003} ($GF(g)=GF(2^q)$). \acp{gf} are constructed  under a specific set of properties, ensuring that any operation performed over a \ac{gf} returns a result within the same field. In computational terms, operations in \acp{gf} allow the simplification of \acp{alu} since circuits are no longer required for carry flags and integer overflow~\cite{Chen:2017}.

Irreducible primitive polynomials define a $g$-order \ac{gf}. \Ac{nbldpc} codes use the finite field in $GF(2^q)$, where the elements are represented as $0, 1, \alpha , \alpha ^2, ..., \alpha ^{2^{q}-2}$, with $\alpha $ being a primitive element of $GF(2^q)$~\cite{carrasco:2008}.

The \ac{nbldpc} code presented in (\ref*{eq:NBPCM}), has the primitive polynomial $\alpha ^2+x^1+1$. \ac{nbldpc} codes compute symbols in their respective \ac{gf}, with each symbol representing $q$ bits. For the example provided in Fig.~\ref*{fig:tanner_graph} ($GF(2^2)$), each symbol represents two bits.

\begin{equation} \label{eq:NBPCM}
  \mathbf{H=\begin{bmatrix}
  \alpha  &0  &1  &\alpha  &0  &1 \\ 
  \alpha ^{2} &\alpha   &0  &1  &1  &0 \\ 
  0 &\alpha   &\alpha ^{2}  &0  &\alpha ^{2}  &1 
  \end{bmatrix} }
  \end{equation}



Formally, a \ac{nbldpc} code size is denoted as $(N, K)$, where $K=N-M$. During transmission, a codeword can be corrupted by noise. At the receiver, the non-binary decoder should receive a codeword $\mathbf{y}$ with $N$ symbols that will be corrected, if necessary.



\subsection{Non-binary LDPC Decoding Algorithms}

In \ac{nbldpc} codes, instead of using single bits ($GF(2)=\left\{0,1\right\} $), the messages are encoded into symbols composed of $q$-tuples of bits (e.g. $GF(2^2)=\left\{00,01,10,11\right\} $) to correct transmission errors over the channel. 

The first step is to compute the probability vector using the data received from the channel for every symbol in $GF(g)$ ($\mathbf{u_{m,n}^{(0)}}(x)=p\left(c_{n}=x|y_{n}\right), x \in GF(2^q)$). Assuming \ac{bpsk} modulation, an \ac{awgn} channel, and that the received bits are independent, the probability of a received symbol can be calculated  through the product of the independent bit. For instance, if the probability  of a received bit being $1$ is $f^{(1)}_{n'}=p\left(c_{n'}=1|y_{n'}\right)$ and being $0$ is $f^{(0)}_{n'}=1-f^{(1)}_{n'}$, the probability  of a symbol being '$11$' in $GF(2^2)$ is $p\left(c_{n}=11|y_{n}\right) = f^{(1)}_{n'}f^{(1)}_{{n'}+1}$. Generalizing for \ac{awgn}, the probability  for each symbol can be calculated  through $p\left(c_{n}=x|y_{n}\right) \propto \prod\limits_{i=0}^{q-1}\frac{1}{1+e^{\frac{2y_{i}b_i}{\sigma^2}}}, b_i=\left\{-1,1\right\}$ and $x \in GF(2^q)$, where the probabilities of the symbols are proportional to the productory, since it requires them to be normalized to guarantee the axioms ($ \sum\limits_{n}p\left(c_{n}=x|y_{n}\right)=1$)~\cite{carrasco:2008}.

\begin{algorithm}[!t]
  \caption{\bf \protect\ac{fft}-\protect\ac{spa} \protect\ac{nbldpc} decoding algorithm}\label{alg_FFT_SPA}
  {\small{\bf Input:}  $\mathbf{H}$; $I_{max}$, $\mathbf{y}$, $\mathbf{\widehat{c}}$ \\
  {\bf Initialization:} $\mathbf{u^{(0)}_{m,n}}(x)=\underset{x \in GF(2^q)}{p\left(c_n=x|y_n\right) }$\\
  \indent\hspace{4em}$\mathbf{p_{n}}(x)=\mathbf{u^{(0)}_{m,n}}(x)$\\
  \bf for $k=1:I_{max}$\\ 
  \indent\hspace{1em}{\bf  Compute $\mathbf{\widehat{c}}^{(k-1)} \mathbf{H}^T$; Stop if $\mathbf{\widehat{c}}^{(k-1)} \mathbf{H}^T=0$}\\
  \indent\hspace{1em}{\bf Permutation}\\
  \indent\hspace{1em}For each \ac{vn} $n \in N$, each $m\in S_c(n)$, each $x\in GF(2^q)$\\
  \begin{equation}
   \label{eq:FFT_Perm}
  \mathbf{u'^{(k)}_{m,n}}(x)=h_{m,n}^{-1}\cdot\mathbf{u^{(k-1)}_{m,n}}(x) \\  
  \end{equation}
  \indent\hspace{1em}{\bf \protect\Acf{cnp}}\\
  \indent\hspace{1em}For each \ac{cn} $m \in M$, each $n\in S_v(m)$, each $x\in GF(2^q)$
  \begin{equation}
  \label{eq:FFT_FFT}
  \mathbf{v^{(k)}_{m,n}}(x)=\mathcal{FFT}^{-1}  \left (   \prod_{j\in S_v(m)\backslash n}\mathcal{FFT} \left (  \mathbf{u'^{(k)}_{m,j}}(x) \right )  \right ) 
  \end{equation}
  \indent\hspace{1em}{\bf Depermutation}\\
  \indent\hspace{1em}For each \ac{cn} $m \in M$, each $n\in S_v(m)$, each $x\in GF(2^q)$\\
  \begin{equation}
  \label{eq:FFT_deperm}
  \mathbf{v'^{(k)}_{m,n}}(x)=h_{m,n}\cdot\mathbf{v^{(k)}_{m,n}}(x^{-1})  \\ 
  \end{equation}
  \indent\hspace{1em}{\bf \protect\Acf{vnp}}\\
  \indent\hspace{1em}For each \ac{vn} $n \in N$, each $m\in S_c(n)$, each $x\in GF(2^q)$\\
  \begin{equation}
\label{eq:FFT_VNP_prod}
  \mathbf{z^{(k)}_{m,n}}(x)=p_n(x)\prod_{i\in S_c(n)\backslash m}v'^{(k)}_{i,n}(x) 
  \end{equation}
  \begin{equation}
    \label{eq:FFT_VNP_Norm}
    \mathbf{u^{(k)}_{m,n}}(x)= \frac{z^{(k)}_{m,n}(x)}{\sum\limits_{i\in  GF(2^q)}z^{(k)}_{m,n}(i)}
  \end{equation}
  \indent\hspace{1em}{\bf A posteriori information computation \& Hard decision}\\
  \indent\hspace{1em}For each \ac{vn} $n \in N$, each $x\in GF(2^q)$\\
    \begin{equation*}
    z'^{(k)}_n(x)=p_n(x)\prod_{i\in S_c(n)}u^{(k)}_{i,n}(x)
    \end{equation*}
    \begin{equation*}
      \widehat{c}^{(k)}_n=\arg\max_{x \in GF(2^q)}\left(\frac{z'^{(k)}_{n}(x)}{\sum\limits_{i\in GF(2^q)}z'^{(k)}_{n}(i)}\right) 
      \end{equation*}
  
  }
  \end{algorithm}

The \ac{nbldpc} decoding algorithm tries to seek a codeword $\mathbf{\widehat{c}}$ such that maximizes $p\left( \mathbf{\widehat{c}}|\mathbf{y}, \mathbf{\widehat{c}}\mathbf{H}^T=0\right)$, meaning that  $\mathbf{\widehat{c}}$ is calculated through an iterative process until satisfying all parity check equations or until a maximum number of iterations ($I_{max}$) is reached. This can be achieved by calculating the maximum \ac{app} of every \ac{vn} and \ac{cn}. However, the calculation of the \acp{app} of every node is computationally expensive and can be simplified by excluding the information of unconnected nodes and considering the set of connected nodes $S_f$ except one node $i$ ($p\left( \widehat{c}_n=x|\mathbf{y}, S_f\backslash i \right)$ with $x \in GF(2^q)$). The node $i$ should cycle through the set $S_f$, allowing the algorithm to calculate the most probable symbols for the decoded codeword $\mathbf{\widehat{c}}$.

In summary, the \ac{nbldpc} decoders exchange messages between \acp{cn} and \acp{vn}, where these messages are vectors containing the probability of each symbol being correct that update when exchanged between nodes

 Davey and MacKay proposed the non-binary formulation of the \ac{spa}~\cite{Davey:1998}, but it has high computational complexity. One solution to decrease complexity in the \ac{cnp} is to avoid the convolution of probabilities and compute them in the frequency domain, as depicted in Algorithm~\ref{alg_FFT_SPA}.

  \begin{algorithm}[!t]
    \caption{\bf \protect\Acf{mm} \protect\ac{nbldpc} decoding algorithm}\label{alg_MM}
    {\small{\bf Input:}  $\mathbf{H}$; $I_{max}$, $\mathbf{y}$, $\mathbf{\widehat{c}}$ \\
    {\bf Initialization:} $p\left(c_n=\eta\right) =\max\limits_{x \in GF(2^q)}\left(p\left(c_n=x|y_n\right)\right)$\\
    \indent\hspace{4em}$\mathbf{\gamma_{n}}(x)=ln\left(\frac{p\left(c_n=\eta \right) }{p\left(c_n=x\right)} \right), x \in GF(2^q)\backslash \eta $\\
    \indent\hspace{4em}$ \alpha_{m,n}^{(0)}(x)=\gamma_{n}(x)$\\
    \bf For $k=1:I_{max}$\\ 
    \indent\hspace{1em}{\bf  Compute $\mathbf{\widehat{c}}^{(k-1)} \mathbf{H}^T$; Stop if $\mathbf{\widehat{c}}^{(k-1)} \mathbf{H}^T=0$}\\
    \indent\hspace{1em}{\bf \protect\Acf{cnp}, F \& B matrices}\\
    \indent\hspace{1em}For each \ac{cn} $m \in M$, each $n\in S_v(m)$, each $x \in GF(2^q)$\\
    \indent\hspace{4em}If $n==0$\\
    \begin{equation} 
    \label{eq:MM_F_first}
      F_{m,n}^{(k)}(x)=\alpha_{m,n}^{(k-1)}\left(h_{m,n}^{-1}\cdot x\right) 
    \end{equation} 
    \begin{equation} 
    \label{eq:MM_B_first}
      B_{m,n+d_c-1}^{(k)}(x)=\alpha_{m,n+d_c-1}^{(k-1)}\left(h_{m,n+d_c-1}^{-1}\cdot x \right) 
    \end{equation}
    \indent\hspace{4em}Else\\
    \indent\hspace{6em}For each $x'\in GF(2^q)$\\ 
    \begin{equation} 
    \label{eq:MM_F_remaining}
      F_{m,n}^{(k)}(x)=\min\limits_{x'+h_{m,n}\cdot x''=x} \left( \max \left(F_{m,n-1}^{(k)}(x'),\alpha_{m,n}^{(k-1)}(x'')\right)  \right) 
    \end{equation} 
    \begin{equation} 
    \label{eq:MM_B_remaining}
      B_{m,n-1}^{(k)}(x)=\min\limits_{x'+h_{m,n-1}\cdot x''=x}\left(\max \left(B_{m,n}^{(k)}(x'),\alpha_{m,n-1}^{(k-1)}(x'')\right)  \right)  
    \end{equation}   
    \indent\hspace{1em}{\bf \protect\Acf{cnp}, $\beta$ matrix}\\
    \indent\hspace{1em}For each \ac{cn} $m \in M$, each $n\in S_v(m)$, each $x \in GF(2^q)$\\
    \indent\hspace{4em}If $n==0$\\
    \begin{equation}
    \label{eq:MM_beta_first}
      \beta_{m,n}^{(k)}(x)=B_{m,1}^{(k)}(h_{m,n}\cdot x)
    \end{equation} 
    \indent\hspace{4em}Else if $n==d_c-1$\\
    \begin{equation} 
    \label{eq:MM_beta_last}
      \beta_{m,n}^{(k)}(x)=F_{m,n-1}^{(k)}(h_{m,n}\cdot x)  
    \end{equation} 
    \indent\hspace{4em}Else\\ 
    \indent\hspace{6em}For each $x'\in GF(2^q)$\\  
    \begin{equation} 
    \label{eq:MM_beta_remaining}
      \beta_{m,n}^{(k)}(x)=\min\limits_{x'+h_{m,n}\cdot x=x''} \left( \max \left(F_{m,n-1}^{(k)}(x''),B_{m,n+1}^{(k)}(x')\right)  \right) 
    \end{equation}  
    \indent\hspace{1em}{\bf \protect\Acf{vnp}}\\
    \indent\hspace{1em}For each \ac{vn} $n \in N$, each $m\in S_c(n)$, each $x\in GF(2^q)$\\
    \begin{equation}
    \label{eq:MM_vnp_sum}
    \alpha'^{(k)}_{m,n}(x)=\gamma_n(x)+\sum_{i\in S_c(n)\backslash m}\beta_{i,n}^{(k)}(x) 
    \end{equation}
    \begin{equation}
    \label{eq:MM_vnp_norm}
    \alpha_{m,n}^{(k)}(x)=\alpha'^{(k)}_{m,n}(x)-\min\limits_{x \in GF(2^q)}\alpha'^{(k)}_{m,n}(x)
    \end{equation}
    \indent\hspace{1em}{\bf A posteriori information computation \& Hard decision}\\
    \indent\hspace{1em}For each \ac{vn} $n \in N$, each $x\in GF(2^q)$\\
      \begin{equation*}
      \widehat{c}^{(k)}_n=\arg\min_{x \in GF(2^q)}\left(\gamma_n(x)+\sum_{i\in S_c(n)}\beta^{(k)}_{i,n}(x)\right) 
      \end{equation*}
    }
    \end{algorithm}

  The \ac{fft}-\ac{spa} reduces the complexity of the \ac{cnp}. However, it requires floating-point (or quantization schemes) and division operators. The decoder complexity can be further reduced by computing the probabilities in the logarithmic domain with respect to the most probable symbol. Savin~\cite{Savin:2008} proposed the \ac{mm} described in Algorithm~\ref{alg_MM}, showing that by using appropriate metrics, the number of complex operations can be reduced, producing a quasi-optimal decoder.

  Furthermore, the messages exchanged in the \ac{fft}-\ac{spa} must be multiplied by $h_{m,n}$ when exchanged from \acp{cn} to \acp{vn} and divided when exchanged from \acp{vn} to \acp{cn} in their respective \acp{gf}. For example, the probabilities are multiplied as follows:

\begin{equation*}
   \mathbf{v'_{m,n}}(x)= \begin{bmatrix}
    v_{m,n}(0)\\ 
    v_{m,n}(1)\\ 
    v_{m,n}(\alpha)\\ 
    \vdots \\ 
    v_{m,n}(\alpha^{2^{q}-2})
    \end{bmatrix},
 \mathbf{v'_{m,n}}(x\cdot \alpha)= \begin{bmatrix}
      v_{m,n}(0)\\ 
      v_{m,n}(\alpha^{2^{q}-2})\\ 
      v_{m,n}(1)\\ 
      v_{m,n}(\alpha)\\
      \vdots 
      \end{bmatrix},
    \end{equation*}

    \begin{equation*}
       \mathbf{v'_{m,n}}(x\cdot \alpha^2)= \begin{bmatrix}
      v_{m,n}(0)\\ 
      v_{m,n}(\alpha^{2^{q}-3})\\ 
      v_{m,n}(\alpha^{2^{q}-2})\\ 
      v_{m,n}(1)\\
      \vdots 
      \end{bmatrix},
    \mathbf{u'_{m,n}}(x\cdot \alpha^{-1})= \begin{bmatrix}
      u_{m,n}(0)\\ 
      u_{m,n}(\alpha)\\ 
      \vdots\\ 
      u_{m,n}(\alpha^{2^{q}-2})\\
      u_{m,n}(1) 
      \end{bmatrix}.
  \end{equation*} 


  \noindent This operation can be performed by executing a barrel shift where the first element is fixed. For instance, the message propagated from $VN_0$ to $CN_0$ in Fig.~\ref{fig:tanner_graph}, will be multiplied by $\alpha$. Some decoding algorithms (such as the \ac{mm}), feature additions in \ac{gf} that can be simplified to an XOR operation.

\begin{table*}[!t]
  \begin{center}

  \caption{Comparison of the number of operations between the \ac{fft}-\ac{spa} and the \ac{mm} decoding algorithms. The presented values represent one decoding iteration. Initialization, a posteriori information, and hard decision are not considered. Bitwise operations (AND, XOR operations) are considered as additions.} 
  \label{table:ops}
  \resizebox{\linewidth}{!}{\begin{tabular}{|ll|l|l|l|l|l|}
\hline
\multicolumn{2}{|l|}{} & \begin{tabular}[c]{@{}l@{}}Additions \& Subtractions\end{tabular} & Multiplications & Divisions & Comparisons & Memory Transactions \\ \hline
\multicolumn{1}{|l|}{\multirow{10}{*}{FFT-SPA}} & \begin{tabular}[c]{@{}l@{}}Permutation\\ Loop Control\end{tabular} & $N+Nd_v+Nd_vg$ & $0$ & $0$ & $N+Nd_v+Nd_vg$  & $0$ \\ \cline{2-7} 
\multicolumn{1}{|l|}{} & Permutation & $0$ & $0$ & $0$ & $0$ & $3Nd_vg$ \\ \cline{2-7} 
\multicolumn{1}{|l|}{} & \begin{tabular}[c]{@{}l@{}}\ac{cnp}\\ Loop Control\end{tabular} & $M+Md_c+Md_cg$ & $0$ & $0$ & $M+Md_c+Md_cg$ & $0$ \\ \cline{2-7} 
\multicolumn{1}{|l|}{} & \ac{fft} & $Md_c g\log_2 g$ & $Md_c g\log_2 g$ & $0$ & $0$ & $Md_c g\log_2 g$ \\ \cline{2-7} 
\multicolumn{1}{|l|}{} & \ac{cnp} Productory & $Md_c(d_c-1)g$ & $Md_c(d_c-1)g$ & $0$ & $0$ & $Md_c(d_c-1)g$ \\ \cline{2-7} 
\multicolumn{1}{|l|}{} & Inverse \ac{fft} & $Md_cg\log_2 g$ & $Md_cg\log_2 g$ & $Md_cg$ & $0$ & $Md_cg\log_2 g+Md_cg$ \\ \cline{2-7} 
\multicolumn{1}{|l|}{} & \begin{tabular}[c]{@{}l@{}}Depermutation\\ Loop Control\end{tabular} & $M+Md_c+Md_cg$ & $0$ & $0$ &$M+Md_c+Md_cg$ & $0$ \\ \cline{2-7} 
\multicolumn{1}{|l|}{} & Depermutation & $0$ & $0$ & $0$ & $0$ & $3Md_cg$ \\ \cline{2-7} 
\multicolumn{1}{|l|}{} & \begin{tabular}[c]{@{}l@{}}VNP\\ Loop Control\end{tabular} & $N+Nd_v+Nd_vg$ & $0$ & $0$ & $N+Nd_v+Nd_vg$ & $0$ \\ \cline{2-7} 
\multicolumn{1}{|l|}{} & VNP & $Nd_vg$ & $Nd_v(d_v-1)g+Nd_vg$ & $Nd_vg$ & $0$ & $5Nd_vg+Nd_v(d_v-1)g$ \\ \hline
\multicolumn{1}{|l|}{\multirow{10}{*}{Min-Max}} & \begin{tabular}[c]{@{}l@{}}F\&B Matrices\\ First Edge\\Loop Control\end{tabular} & $M+Md_c+Md_cg$ & $0$ & $0$ & $M+Md_c+Md_cg$ & $0$ \\ \cline{2-7} 
\multicolumn{1}{|l|}{} & \begin{tabular}[c]{@{}l@{}}F\&B Matrices\\ First Edge\end{tabular} & $0$ & $0$ & $0$ & $Md_cg$ & $6Mg$ \\ \cline{2-7} 
\multicolumn{1}{|l|}{} & \begin{tabular}[c]{@{}l@{}}F\&B Matrices\\ Remaining Edges\\ Loop Control\end{tabular} & $M(d_c-1)g^2$ & $0$ & $0$ & $M(d_c-1)g^2$ & $0$ \\ \cline{2-7} 
\multicolumn{1}{|l|}{} & \begin{tabular}[c]{@{}l@{}}F\&B Matrices\\ Remaining Edges\end{tabular} & $2M(d_c-1)g^2$ & $0$ & $0$ & $2M(d_c-1)g^2+2M(d_c-1)g$ & $8M(d_c-1)g^2$ \\ \cline{2-7} 
\multicolumn{1}{|l|}{} & \begin{tabular}[c]{@{}l@{}}$\beta$ Matrix\\ First and Last Edge\\Loop Control\end{tabular} & $M+Md_c+Md_cg$ & $0$ & $0$ & $M+Md_c+Md_cg$ & $0$ \\ \cline{2-7} 
\multicolumn{1}{|l|}{} &\begin{tabular}[c]{@{}l@{}}$\beta$ Matrix\\ First and Last Edge\end{tabular} & $0$ & $0$ & $0$ & $2Md_cg$ & $6Mg$ \\ \cline{2-7} 
\multicolumn{1}{|l|}{} & \begin{tabular}[c]{@{}l@{}}$\beta$ Matrix\\ Remaining Edge\\Loop Control\end{tabular} & $M(d_c-2)g^2$ & $0$ & $0$ & $M(d_c-2)g^2$ & $0$ \\ \cline{2-7} 
\multicolumn{1}{|l|}{} & \begin{tabular}[c]{@{}l@{}}$\beta$ Matrix\\ Remaining Edge\end{tabular} & $M(d_c-2)g^2$ & $0$ & $0$ & $M(d_c-2)g^2+M(d_c-2)g$ & $4M(d_c-2)g^2$ \\ \cline{2-7} 
\multicolumn{1}{|l|}{} &  \begin{tabular}[c]{@{}l@{}} VNP\\ Loop Control\end{tabular} & $N+Nd_v+Nd_vg$ & $0$ & $0$ & $N+Nd_v+Nd_vg$ & $0$ \\ \cline{2-7} 
\multicolumn{1}{|l|}{} & VNP & $Nd_v(d_v-1)g+2Nd_vg$ & $0$ & $0$ & $Nd_vg$ & $5Nd_vg+Nd_v(d_v-1)g$ \\ \hline
\end{tabular}}
  \end{center}
  \end{table*}

  The \ac{mm} decoding algorithm computes the \ac{llr} vectors with respect to the most likely symbol ($LLR=ln\left(\frac{p\left(c_n=\eta _n|y_n\right) }{p\left(c_n=x|y_n\right)} \right) x \in GF(2^q) $), resulting in a vector with $2^q-1$ elements. The most likely symbol of the field can calculated through $p\left(c_n=\eta_n\right) =  \max\left( p\left(c_n=x|y_n\right)\right), x \in GF(2^q)  $.

  The main difference lies in the \acp{cnp}, which replace sums and products by comparisons, using a forward and backward computation technique~\cite{Savin:2008}. Furthermore, using \acp{llr} instead of the most likely  symbol, reduces  the number of comparisons between symbols in the same message, reducing computational complexity.

\subsection{Theoretical Decoding Complexity}


The number of operations and memory transactions of each decoding algorithm are presented in Table~\ref{table:ops}. Additions/subtractions in \acp{gf} are equivalent to \textit{XOR} operations and multiplications/divisions in \acp{gf} are equivalent to read and write operations (barrel shift operations). In order to improve performance, these operations are implemented in \acp{lut}.

The \ac{fft}-\ac{spa} uses a high number of additions and multiplications in the \ac{cnp} (complexity of $\mathcal{O} \left(Md_cg\log_2g\right)$) and \ac{vnp} (complexity of $\mathcal{O} \left(Nd_v^2g\right)$), where the \ac{cnp} complexity dominates the the complexity of the \ac{fft}-\ac{spa}, as compared to the \ac{vnp}. The overall expression representing the total number of operations (excluding memory transactions) is $4M+4Md_c+3Md_cg+2Md_c^2g+4Md_cg\log_2g+4N+4Nd_v+6Nd_vg+Nd_v^2g$.

In terms of memory transactions, the complexity is mostly located in the \ac{fft}, inverse \ac{fft}, and \ac{vnp}. The memory transaction complexity in the \ac{cnp} is $\mathcal{O} \left(Md_cg\log_2g\right)$ and for the \ac{vnp} is $\mathcal{O} \left(Nd_v^2g\right)$, where the overall complexity is the same as the complexity of the \ac{cnp} ($M\gg d_v^2$). The total number of memory transactions in the \ac{fft}-\ac{spa} is described by $3Md_cg+Md_c^2g+2Md_cg\log_2g+7Nd_vg+Nd_v^2g$.

In comparison to the \ac{fft}-\ac{spa}, the \ac{mm} algorithm transforms multiplications and divisions into additions/subtractions and comparisons, with most operations located in the Forward and Backward, and $\beta$ matrices computation. The \ac{cnp} has a computational complexity of $\mathcal{O} \left(Md_cg^2\right)$, while the \ac{vnp} has a complexity of $\mathcal{O} \left(Nd_v^2g\right)$, with an overall computational complexity $\mathcal{O} \left(Md_cg^2\right)$. The general expression describing the total number of operations in \ac{mm} without memory transactions is $4M+4Md_c-4Mg-14Mg^2+10Md_cg+10Md_cg^2+2N+2Nd_v+4Nd_vg+Nd_v^2g$. 

 The Forward and Backward, and $\beta$ matrices have the highest number of memory transactions in the \ac{mm}. The memory transaction complexity  is the same as the computational complexity, with the expression $12Mg-16Mg^2+12Md_cg^2+Nd_v^2g+4Nd_vg$ representing the total number of memory transactions in the \ac{mm}.

The theoretical computational and memory transaction complexity mainly scales with the size of \ac{gf}, since $d_c$ is usually lower than $27$~\cite{Ferraz:2021}. At first glance, \ac{pim} solutions might not seem attractive to implement \ac{ldpc} decoders due to their limited clock speed and \acp{alu} compared to other devices. However, by using reduced precision formats and several thousands of decoders in parallel, \ac{pim} systems can counteract the downsides of having lower clock frequencies.

\section{UPMEM System}



Parallel processing platforms, such as \acp{gpu}, \acp{fpga}, and \acp{asic}, can offload specific tasks from the \ac{cpu} to further enhance computational capabilities by allowing multiple processes to run simultaneously. These platforms leverage concurrency to solve problems more quickly, either by dividing tasks into smaller subtasks that can be processed in parallel or by executing different tasks in parallel to increase throughput. 

Programming models for \acp{cpu}, \acp{gpu}, \acp{fpga}, and \acp{asic} differ significantly due to their distinct architectures and intended applications. For example, \acp{cpu} use programming models optimized for task scheduling and execution on operating systems. On the other hand, \acp{gpu} are designed for massive parallelism, using thousands of lightweight threads that execute the same instructions on different data.

The UPMEM system is a massive multicore-\ac{cpu} with private memory that employs several simple on-chip computing units using a \ac{risc}-based \ac{isa}. Using memory and processing units on the same chip imposes challenges in manufacturing design. For instance, memory technology is optimized for density (more memory cells per unit of area or volume), and processor technology is optimized for speed. Combining the two technologies causes thermal dissipation problems, causing the processing unit's clock frequency to be reduced, the reduction of memory density, and the employment of active cooling strategies to increase the system's complexity with limited effectiveness (memory technology still needs low temperatures to operate).

UPMEM addresses these problems by implementing near-memory processing units, called \acfp{dpu},  with relatively deep pipelines and fine-grained multithreading running at several hundred megahertz.

\subsection{UPMEM Architecture}  
\label{sec:UPMEM_arch}

\begin{figure}[!t] 
  \centering
  \includegraphics[width=0.5\columnwidth]{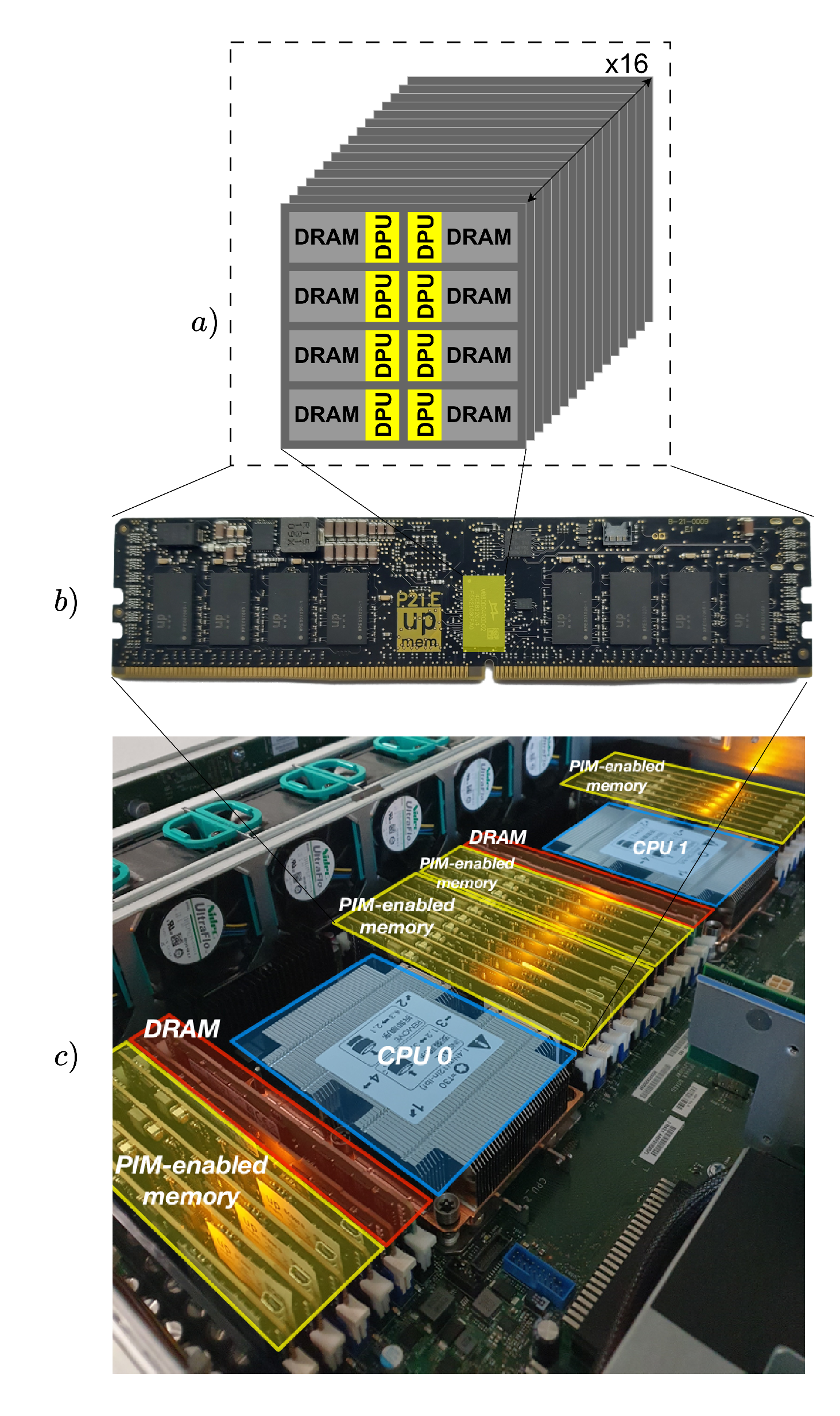}
  \caption{UPMEM system strucutre. $a)$ \ac{pim} chip is composed of eight \acp{dpu} and respective \ac{dram} banks. Each \ac{pim} \ac{dimm} comprises $16$ \ac{pim} chips. $b)$ includes UPMEM \ac{pim} module  $16$ \ac{pim} chips, eight at the front and eight at the back. Each \ac{pim} chip is composed of eight \acp{dpu}. $c)$ UPMEM \ac{pim} server contains two \acp{cpu}, $20$ \ac{pim} modules, and four conventional \ac{dram} modules. Courtesy of~\cite{Devaux:2019}  }
  \label{fig:UPMEM}
\end{figure}

\begin{figure}[!t]
  \centering
  \includegraphics[width=\columnwidth]{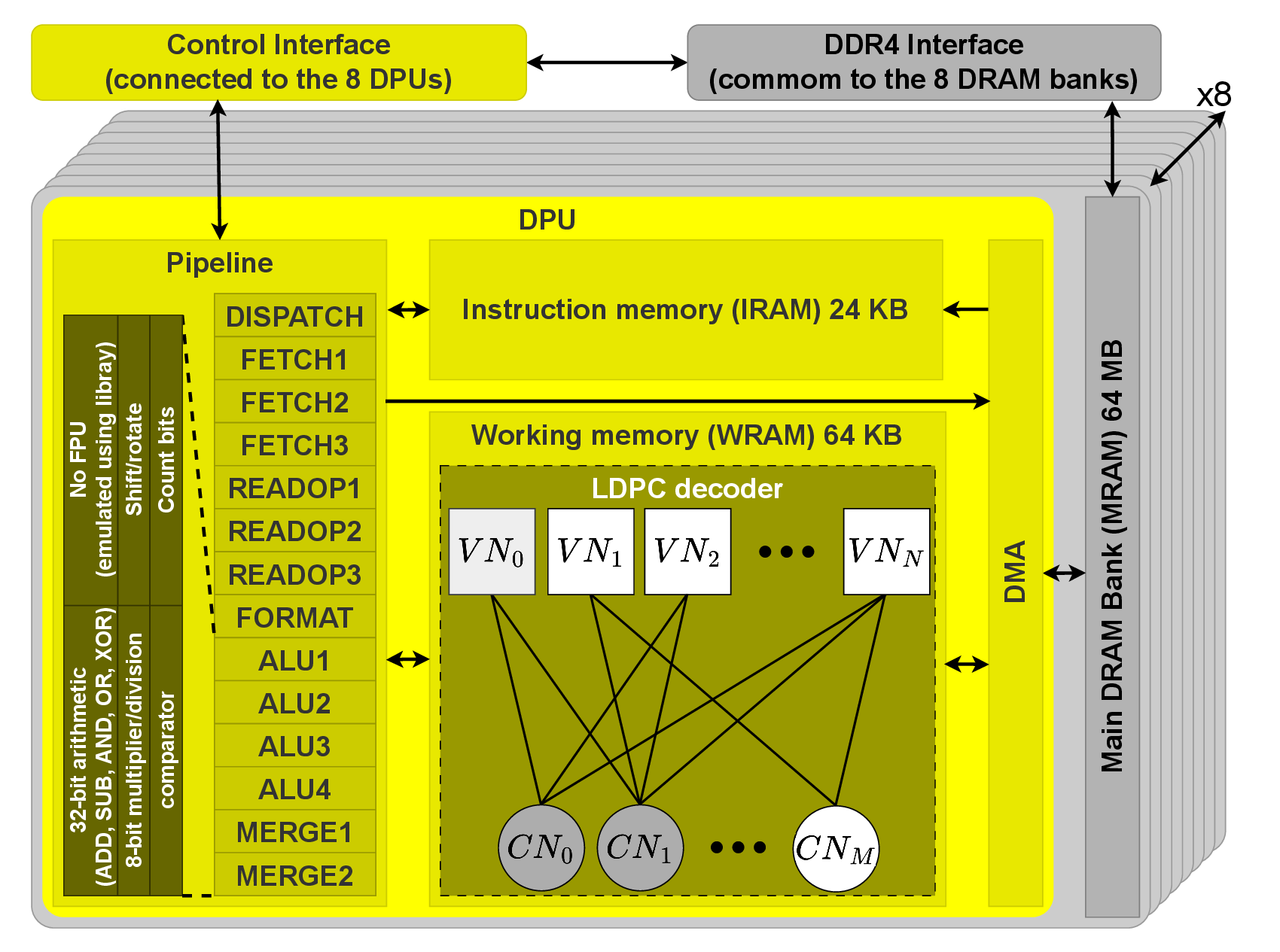}
  \caption{Architecture of a UPMEM \ac{pim} chip for \ac{ldpc} decoding. Each \ac{pim} chip comprises eight \acp{dpu} and independent \ac{mram} banks. Each \ac{dpu} contains  $24$ KB \ac{iram} and a $64$ KB \ac{sram}-based scratchpad memory (\ac{wram}). The \ac{dpu} has a $14$-stage pipeline,  supports $32$-bit arithmetic, $8$-bit multipliers/division, emulates $32$-bit multiplications, and \ac{fp} operations are emulated using IEEE-$754$ software libraries. The variables, topology, and codewords of the \ac{ldpc} decoder are saved in \ac{wram} and off-loaded to \ac{mram} when no space is available. Several decoder are saved in \ac{wram}.}
  \label{fig:DPU}
\end{figure}

A general overview of the UPMEM system is shown in Fig.~\ref{fig:UPMEM}, and the internal architecture of a single \ac{dpu} is detailed in Fig.~\ref{fig:DPU}. A \ac{dpu} is a $32$-bit \ac{risc}-based unit that operates in-order and supports multithreading with a proprietary \ac{isa}~\cite{Devaux:2019}. Each \ac{dpu} supports up to $24$ hardware threads, each with $24$ $3$2-bit  registers. As depicted in Fig.~\ref{fig:DPU}, \ac{dpu} has a $14$-stage pipeline. However, only the last three stages (ALU4, MERGE1, and MERGE2) are executed in parallel with the DISPATCH and FETCH stages within the same thread. Therefore, instructions from the same thread are dispatched in $11$-cycle intervals, which requires only $11$ threads to fully exploit the pipeline~\cite{gomez:2021,Devaux:2019}.

The \ac{dpu} \ac{alu} supports $32$-bit additions/subtractions and bitwise operations, shift/rotate operations, bit counters, and $8$-bit multiplications/divisions. $32$-bit multiplications and divisions are not natively supported due to the high hardware cost and the limited number of metal layers~\cite{Devaux:2019}, but are emulated by using $8$-bit multiplications/divisions and shift operations. Furthermore, the \ac{dpu} does not have \ac{fp} functional units and emulates \ac{fp} operations using software libraries ($\sim 200$ cycles for a $32$x$32$ \ac{fp} multiplication).

The hardware threads use a shared instruction memory (\ac{iram})  and a scratchpad memory (\ac{wram}). The $24$ KB \ac{iram} supports up to $4096$ $48$-bit encoded instructions. The faster \ac{sram}-based \ac{wram} has a size of $64$ KB and can be accessed through $8$, $16$, $32$, and $64$-bit load/store operations. The \acp{dpu} also include a slower $64$ MB \ac{dram}-based \ac{mram}, which uses \ac{dma} instructions to transfer data from the \ac{mram} bank to the \ac{iram} and \ac{wram}.

Each UPMEM \ac{pim} chip is composed of eight \acp{dpu} with a control and a \ac{ddr4} interfaces. Each UPMEM module comprises $16$ \ac{pim} chips, as shown in Fig.~\ref{fig:UPMEM}. The UPMEM server includes two \acp{cpu}, conventional \ac{dram} modules, and $20$ \ac{pim} modules, as shown in Fig.~\ref{fig:UPMEM} $c)$ containing a total of $2560$ \acp{dpu} and $160$ GB of \ac{mram} running at $350$ MHz.

\subsection{UPMEM Programming Model}  

The UPMEM follows a \ac{spmd} programming model, where the same binary is loaded in each \ac{dpu}, and threads execute the same code on separate data fragments.  The UPMEM \ac{sdk} includes a C LLVM-based compiler, debugger, and runtime and host libraries, allowing the \ac{dpu} programs to be written in the C language and the host program to be written in C, C++, Python, or Java~\cite{UPMEM_manual}.

The host \ac{cpu} compiles both host and device code. The programmer is responsible for allocating the number of \acp{dpu} and threads, defining the data to be transferred between the host and the device, and setting the synchronous/asynchronous kernel execution mode. Synchronous execution stops the \ac{cpu} thread until all \acp{dpu} have finished executing the kernel, while asynchronous execution immediately grants control back to the host \ac{cpu}.

Parallel data transfers from \ac{cpu} to \ac{dpu} can be performed using three different functions: $1)$ \texttt{dpu\_copy\_to()} copies data to a single-\ac{dpu}. $2)$ \texttt{dpu\_push\_xfer()} copies different chunks of data from a single buffer to a set of \acp{dpu}. $3)$ \texttt{dpu\_broadcast\_to()} copies data from a single buffer to several \acp{dpu} within the same set. Two functions can be used to retrieve data from \acp{dpu}: $1)$ \texttt{dpu\_copy\_from()} copies data from a single-\ac{dpu} to the host buffer. $2)$ \texttt{dpu\_push\_xfer()} copies data from several \acp{dpu} to a single host buffer (the function \texttt{dpu\_push\_xfer()} can copy data from or to \acp{dpu} by setting a function's argument). The transfer sizes are the same, using \texttt{dpu\_push\_xfer()} and \texttt{dpu\_broadcast\_to()} functions results in parallel transfers and increases bandwidth.

Data transfers between \ac{mram} and \ac{wram} do not have cache mechanisms, which the programmer must explicitly manage by using \texttt{mram\_read()} (\ac{mram} to \ac{wram} transfers) and \texttt{mram\_write()} (\ac{wram} to \ac{mram} transfers). 

The UPMEM runtime library contains thread control and synchronization mechanisms, such as mutexes that define critical sections between threads, semaphore counters, barriers that synchronize several threads at a specific point in the code, and handshakes that enable synchronization between pairs of threads.

\subsection{Comparison to GPU Architecture}

From a high-level perspective, the UPMEM system shares some similarities with \acp{gpu}. However, their architectures and hierarchies are fundamentally different.

The biggest difference lies in the memory hierarchy. \Acp{gpu} have a main \ac{dram}, which is accessible by all threads in the \ac{gpu}, and a faster \ac{sram} shared memory, which is accessible by threads in the same streaming processor. UPMEM also has \ac{dram} and \ac{sram} but they are accessible by threads within the same \ac{dpu}. This configuration increases data synchronization costs in the UPMEM system since data transmitted between \acp{dpu} travels through the memory bus.

The second difference is that \acp{dpu} have a simple in-order scalar \ac{alu} with no stall signals and instruction pipelining, allowing them to perform different stages of different threads in the same clock cycle. In contrast, \acp{gpu} have several  vector \acp{alu} grouped in a \ac{sm}. This allows them to perform operations at different data points in the same clock cycle with several dedicated \ac{simd} lanes. Furthermore, \acp{gpu} have schedulers that enable simultaneous multithreading by dynamically issuing instructions to maximize performance.

Both systems use an \ac{spmd} programming model, where the same program is loaded into different processing units (into \acp{dpu} in UPMEM, into \acp{sm} in \acp{gpu}). However, the execution model of \acp{gpu} is different than UPMEM. \acp{gpu} have a \ac{simt} execution model with several independent \ac{simd} functional units, sharing \ac{dram}, \ac{sram}, and cache between elements in the same \ac{sm}, while the UPMEM system has several scalar \acp{alu} with interleaved pipelining, that have \ac{dpu}-visible \ac{dram} and \ac{sram} without caching techniques.

\section{Near-memory LDPC Parallel Implementation}

This work selects two widely used \ac{nbldpc} decoding algorithms: the \ac{fft}-\ac{spa} decoder and the \ac{mm} decoder. The \ac{fft}-\ac{spa} is used to evaluate the decoder performance when using \ac{fp} arithmetic, serving as a performance reference. The \ac{mm} decoder, on the other hand, is implemented using fixed-point arithmetic to explore the low-complexity, quantized decoding regime. This combination allows to assess the trade-offs between \ac{fp} emulation and native integer operations on the UPMEM platform.

The proposed parallelization methods for \ac{ldpc} decoders employ an \ac{spmd} model for several simple processing cores. The implementations for binary \ac{ldpc} decoders in \ac{gpu} and the UPMEM system have been published in~\cite{Ferraz:2023, Ferraz:2024}.

While this work follows the same multicodeword mapping strategy used in prior binary \ac{ldpc} implementations on UPMEM~\cite{Ferraz:2023, Ferraz:2024}, the implementation of \ac{nbldpc} decoders introduces significant differences. \Ac{nbldpc} decoding involves operations over \acp{gf}, which require more complex arithmetic (finite field multiplication, \ac{fft}-based convolutions, \ac{llr} updates) and larger message representations. As a result, the memory footprint and computational requirements are substantially higher compared to binary \ac{ldpc} decoding. To accommodate these demands, the \ac{wram} usage must be optimized carefully, and additional synchronization points are required in multithreaded implementations. Furthermore, the \ac{nbldpc} decoding process includes additional permutation and depermutation steps that are not present in binary decoding.

Two approaches can be taken to implement the \ac{ldpc} decoders. The first multicodeword approach assumes weak scaling by assigning a decoder to each \ac{dpu} and increasing the number of decoders by rising the number of \acp{dpu} selected, thus eliminating the need for inter-\ac{dpu} synchronization. The second approach partitions the workload over several \acp{dpu}, with the cost of inter-\ac{dpu} overhead increasing with the number of used \acp{dpu} and the code characteristics (number of nodes, edges, iterations). In order to eliminate data communication overheads between \acp{dpu} and increase performance (assuming the same number of codewords in both strategies), the first approach is chosen for this work,  analyzing single-thread single-\ac{dpu} and multithread-multiple-\acp{dpu} performances.

The multithreaded approach distributes the load over more than $11$ threads to fully utilize the \ac{dpu} pipeline. The \ac{nbldpc} codes used in this work have sizes that are powers of two. Therefore, $16$ threads were used since it is the first power of two bigger than $11$ (please revisit section~\ref{sec:UPMEM_arch} for an explanation on the $11$ threads choice).

Although high levels of decoder replication are often used to maximize throughput in conventional parallel architectures, in the UPMEM system, each \ac{dpu} executes independently and asynchronously. This allows decoding tasks to begin as soon as codewords become available, without requiring synchronization or batching. While there may be an initial latency overhead due to setup and data transfer, this cost is quickly amortized as more codewords are decoded in parallel. Therefore, the proposed multicodeword solution not only enables high throughput but also maintains low per-codeword latency, making it suitable for real-time and low-latency communication scenarios involving short \ac{nbldpc} codewords.

\subsection{Optimization Techniques}

Several optimization techniques can be implemented to enhance the performance of \ac{ldpc} decoders by exploiting the UPMEM system. These techniques maximize parallelism, minimize communication overheads, and efficiently utilize computational resources.

\textbf{\protect\Ac{wram} usage:} \ac{wram} should be preferred over \ac{mram} by storing the Tanner graph characteristics as depicted in Fig.~\ref{fig:DPU}. An \ac{mram}/\ac{wram} read and write takes between $61$ and $1085$ cycles (depending on the data transfer size) and can achieve a bandwidth  of $630$ MB/s~\cite{gomez:2021}. On the other  hand, $8$-byte loads and stores in \ac{wram} only take one clock cycle when the pipeline is full and can achieve a bandwidth  of  $2800$ MB/s~\cite{gomez:2021}.

In our implementation, the partitioning between \ac{mram} and \ac{wram} is manually managed by the programmer, as the UPMEM \ac{sdk} does not provide mechanisms for automatic detection of data reuse or dynamic memory placement. All decoder buffers are allocated in \ac{wram} for fields up to $GF(64)$, which allows faster access and higher throughput. For $GF(128)$ and $GF(256)$, the \ac{wram} capacity is insufficient to store all required buffers, and less frequently accessed data, such as auxiliary buffers used in the \ac{fft} or $\beta$ matrix computation, is explicitly offloaded to \ac{mram}.

\textbf{Quantization:} \Ac{ldpc} decoders use an \ac{fp} representation to calculate the probabilities or \acp{llr} of bits in $\mathbb{R}$. However, using \ac{fp} arithmetic is impractical due to area, power, and complexity constraints, specifically in hardware implementations. A workaround for  these constraints is to use quantization schemes and  integer or fixed-point arithmetic, resulting in decoders that are faster, less complex, more power efficient, and with a low memory footprint, at the cost of a small decrease in error-correction capability~\cite{Gal:2016}.

In the UPMEM system, the \acp{dpu} do not feature \ac{fp} \acp{alu}, and these operations are instead emulated in software, taking between tens and thousands of cycles to perform. The maximum throughput of \ac{fp} $32$-bit arithmetic operations is $4.91$ MOPS for additions, $1.91$ MOPS for multiplications, and $0.34$ MOPS for divisions.

The \acp{dpu} have a $32$-bit integer \ac{alu} that can achieve $58.56$ MOPS for additions, $10.27$ MOPS for multiplications, and $11.27$ MOPS for divisions. Multiplications and divisions have lower throughput due to \acp{dpu} having an $8$-bit multiplier and emulating $32$-bit multiplications/divisions using shifting operations, taking up to $32$ cycles. The $8$-bit multiplier performs similarly to the $32$-bit \ac{alu} when executing $8$-bit arithmetic.

To take advantage of the hardware capability of \acp{dpu}, quantization of the data received from the channel ($y_n$) is performed to the nearest integer. A clipping function is also used to ensure that the values are between $-128$ and $127$.

The effect of quantization on error-correction capability has been studied in the literature. In particular, Wymeersch~\textit{et al.}~\cite{wymeersch:2004:Computational} show that using fixed-point quantization in non-binary \ac{lspa} decoding results in minimal degradation in \ac{ber} performance. Their results indicate that quantization introduces a small performance loss, typically below $0.1$ dB without leading to visible error floors, even in high-\ac{snr} regimes. This supports the use of quantization as a practical trade-off for performance in platforms with limited \ac{fp} support such as UPMEM.

\textbf{Loop unrolling:} Loop unrolling removes branching operations from the \ac{alu}'s pipeline. Thus reducing the total number of operations at a cost of a higher memory footprint. The lack of automatic code optimization of the UPMEM toolchain requires the programmer to optimize the code manually. Loop unrolling is used in the \ac{fft} and inverse \ac{fft} operations of the \ac{fft}-\ac{spa}, leading to increased performance. However, for higher \acp{gf}, the unrolled code does not fit in \ac{iram}, and typical loops must be used. 




\subsection{FFT-SPA}

The \ac{fft}-\ac{spa} described in Algorithm~\ref{alg_FFT_SPA}, uses six processing blocks: (\ref{eq:FFT_Perm}) represents the Permutation block, in (\ref{eq:FFT_FFT}) are represented the \ac{fft}, \ac{cnp} productory, and inverse \ac{fft} blocks, the depermutation in (\ref{eq:FFT_deperm}), and the \ac{vnp} block is represented by (\ref{eq:FFT_VNP_prod}) and (\ref{eq:FFT_VNP_Norm}). For the multithreaded implementation, in all blocks, the edges/messages ($Md_c$ or $Nd_v$) are processed by $16$ threads. Due to data dependencies, synchronization barriers are placed between each block to prevent data hazards.

\begin{table}[!t]
  \begin{center}

  \caption{Number of synchronizations for the \ac{nbldpc} \ac{fft}-\ac{spa} multithreaded decoder on the UPMEM system.} 
  \label{table:ops_spa}
  \resizebox{0.3\columnwidth}{!}{\begin{tabular}{|ll|}
\hline
\multicolumn{2}{|c|}{\textbf{FFT-SPA}}                                                \\ \hline
\multicolumn{1}{|c|}{\textbf{Function}}      & \multicolumn{1}{c|}{\textbf{\# of syncs.}} \\ \hline
\multicolumn{1}{|l|}{Permutation}   & $2$                                      \\ \hline
\multicolumn{1}{|l|}{FFT}           & $2\log_2g$                                     \\ \hline
\multicolumn{1}{|l|}{\ac{cnp}}           & $1$                                      \\ \hline
\multicolumn{1}{|l|}{Inverse FFT}   & $2\log_2g+1$                                   \\ \hline
\multicolumn{1}{|l|}{Depermutation} & $2$                                      \\ \hline
\multicolumn{1}{|l|}{\ac{vnp}}           & $2$                                      \\ \hline
\multicolumn{1}{|l|}{Total}         & $4\log_2g+8$                                   \\ \hline
\end{tabular}}
  \end{center}
  \end{table}

\begin{table}[!t]
  \begin{center}

  \caption{Number of synchronizations for the \ac{nbldpc} \ac{mm} multithreaded decoder on the UPMEM system.} 
  \label{table:ops_mm}
  \resizebox{0.3\columnwidth}{!}{\begin{tabular}{|ll|}
\hline
\multicolumn{2}{|c|}{\textbf{Min-Max}} \\ \hline
\multicolumn{1}{|c|}{\textbf{Function}} & \multicolumn{1}{c|}{\textbf{\# of syncs.}} \\ \hline
\multicolumn{1}{|l|}{F \& B Matrices} & $M+M(d_c-1)$ \\ \hline
\multicolumn{1}{|l|}{$\beta$ Matrix} & $2M+M(d_c-2)$ \\ \hline
\multicolumn{1}{|l|}{VNP} & $1$ \\ \hline
\multicolumn{1}{|l|}{Total} & $2Md_c+1$ \\ \hline
\end{tabular}}
  \end{center}
  \end{table}

The Permutation block (\ref{eq:FFT_Perm}) shifts the message using a Galois multiplication \ac{lut}, saving the permutated messages in a temporary buffer to avoid \ac{raw} and with a synchronization barrier before the permutated messages are written back to memory.

The \ac{fft} block employs a radix-$2$ implementation~\cite{andrade:2013}, which converts the convolution into the frequency domain, transforming the convolution operation into products. The outer loop of this block iterates between $2$ and $q$, which is lower than $16$. Using fewer than $11$ threads leads to an underutilization of the pipeline. Therefore, the chosen approach is to parallelize the inner loop (distribute the edges over threads) and place synchronization barriers at the end of each bit iteration. Furthermore, the inner loop has a synchronization barrier to remove \ac{raw} hazards, bringing the total number of synchronizations to $2\log_2g$ in this block, as shown in Table~\ref{table:ops_spa}.

Once in the frequency domain, all the configurations of the messages connected to the same \ac{cn} are multiplied in the \ac{cnp} productory block.

Next, the inverse \ac{fft} block is applied to the messages. This block applies the same operations as the \ac{fft}. In addition, the normalization step in the \ac{fft} block is transferred to this block, where the messages are divided by $2^q$. The messages are then passed to the Depermutation block (\ref{eq:FFT_deperm}), which applies the same operations as the Permutation block.

In the \ac{vnp} block (\ref{eq:FFT_VNP_prod}), the same operations of the \ac{cnp} productory are performed, but instead to all configurations of messages connected to the \acp{vn}. The probability received from the channel is also considered in this operation. After the new messages are calculated and synchronized, they are normalized in  (\ref{eq:FFT_VNP_Norm}) by the sum of all the probabilities in the same message (sum $= 1$) to grant the validity of the probabilites' axioms.

\subsection{Min-Max}

The \ac{mm} implementation described in Algorithm~\ref{alg_MM} comprises two processing blocks: the first \ac{cnp} block is composed of the Forward (eq. (\ref{eq:MM_F_first}) and (\ref{eq:MM_F_remaining})) and Backward (eq. (\ref{eq:MM_B_first}) and (\ref{eq:MM_B_remaining})) matrices computation (F \& B matrices) and the $\beta$ matrix computation (eq. (\ref{eq:MM_beta_first}),  (\ref{eq:MM_beta_last}), and (\ref{eq:MM_beta_remaining})). The second \ac{vnp} block is composed of (\ref{eq:MM_vnp_sum}) and (\ref{eq:MM_vnp_norm}). The implementation follows the same data structure as the \ac{fft}-\ac{spa}, where the parallelism is exposed to the edges of the \ac{ldpc} code, and the synchronization barriers are placed at the end of each block, as described in Table~\ref{table:ops_mm}. 

The \ac{cnp} block permutates the symbols using the schemes described in Algorithm~\ref{alg_MM}. The F \& B matrices are computed based on the probabilities received from the \acp{vn}. The forward matrix (\ref{eq:MM_F_first}) (\ref{eq:MM_F_remaining}) is computed by comparing the probability from the previous edge and the probability of the received symbol from the connected \ac{vn}. The backward matrix (\ref{eq:MM_B_first}) (\ref{eq:MM_B_remaining}) is calculated in the same way, but instead of considering the probability  from the previous edge, it uses the probability from the next edge. Therefore, the forward matrix is computed from the first to the last edge, and the backward matrix is computed from the last to the first edge. The $\beta$ computation (\ref{eq:MM_beta_first}) (\ref{eq:MM_beta_first}) (\ref{eq:MM_beta_remaining})  compares the permutated probabilities between the Backward matrix's next edge and the Forward matrix's previous edge.

\begin{figure}[t]
  \centering
  \includegraphics[width=0.9\columnwidth]{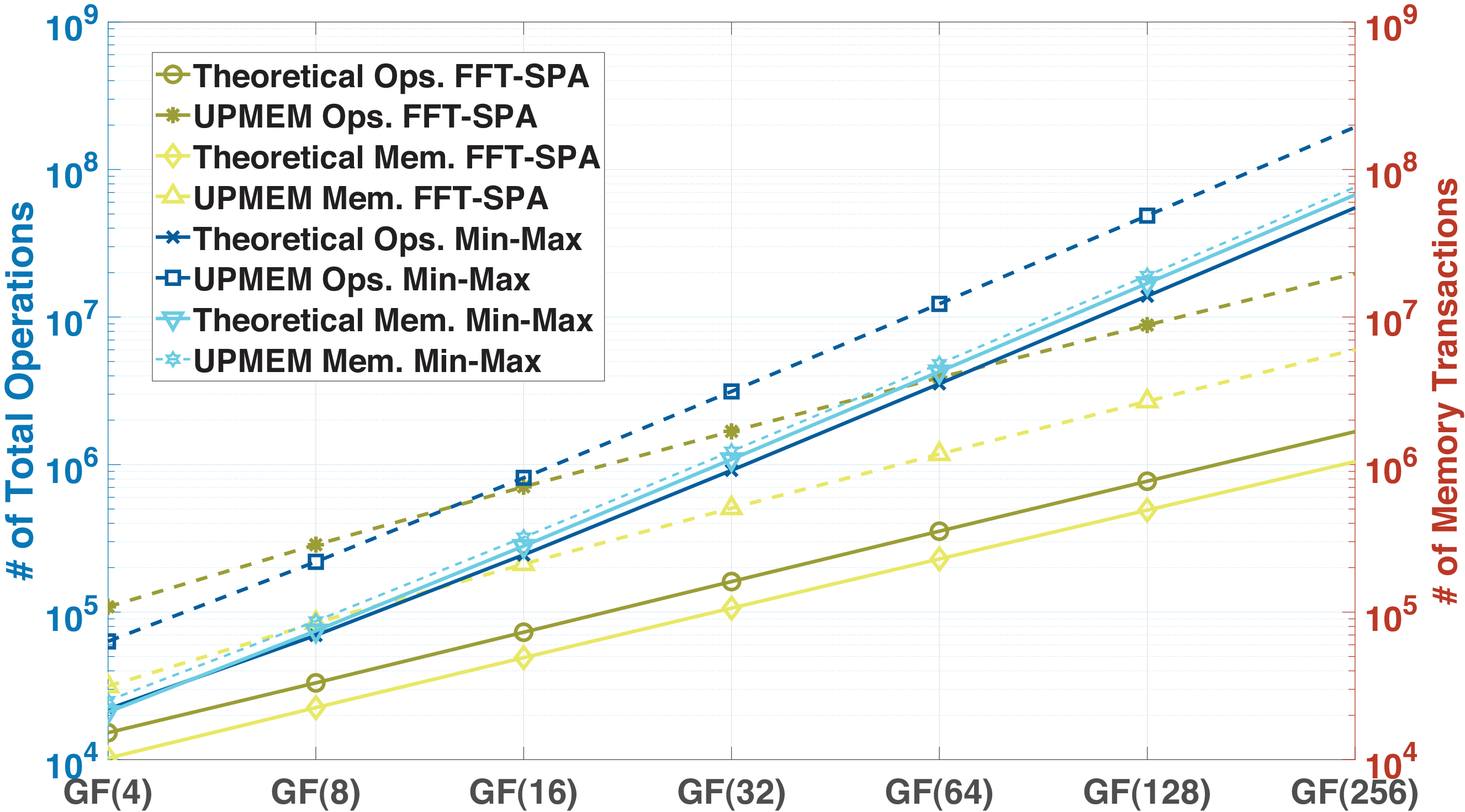}
  \caption{Number of operations and memory transactions to perform one decoding iteration of the $C3$ code. The solid lines represent the theoretical operations while the dashed lines represent counted operations. The yellow lines represent the \ac{fft}-\ac{spa} values while the blue lines depict the \ac{mm}. The vertical axis is represented in the logarithmic scale. The \ac{fft}-\ac{spa} scales better than the \ac{mm} algorithm, but the \ac{mm} requires fewer operations for smaller fields.}
  \label{fig:GF_ops}
\end{figure}


The multithreaded approach exposes parallelism to inner loops and computes several fields in parallel. The limitation of this approach is that for smaller fields ($GF(4)$ and $GF(8)$), it is only possible to use four or eight threads, resulting in a less efficient parallel code due to an under-utilization of the pipeline for lower fields.

The \ac{vnp} block does not have the same issues as the \ac{cnp} block, and the edges/messages are distributed over $16$ threads. This block computes the sum of all configurations of messages ($\beta$ matrix) connected to the same \ac{vn} and adds the sum to the probability received from the channel (\ref{eq:MM_vnp_sum}). To finalize, the newly calculated probabilities are subtracted by the lowest probability (\ref{eq:MM_vnp_norm}) to force the output of the \acp{cn} to have the same \ac{llr} structure as defined in the initialization of the \ac{mm}~\cite{Savin:2008, Declercq:2007}.

\section{Experimental Results}


The \ac{nbldpc} codes used in this work are constructed from a non-binary code for $GF(2^8)$ from~\cite{CCSDS:2015}, resulting in three different \ac{nbldpc} \acp{pcm} of sizes $(16, 8)$, $(32, 16)$, and $(64, 32)$ with $d_c=4$ and $d_v=2$ for $GF(2^2)$ up to $GF(2^8)$. For simplicity, the $(16, 8)$, $(32, 16)$, and $(64, 32)$ codes are designated as $C1$, $C2$, and $C3$ respectively.

The tests were performed on the UPMEM system and compared to low-power \acp{gpu} in the Jetsons Nano $2$ GB, TX$2$, and AGX Xavier, as shown in Table~\ref{table:specs}. We developed all the parallel implementations proposed under the context of this work.

\begin{table*}[t]
  \begin{center}

  \caption{Summary of the platforms specifications used in this work.} 
  \label{table:specs}
  \resizebox{\textwidth}{!}{\begin{tabular}{|l|l|l|l|l|}
\hline
 & UPMEM & Jetson Nano 2 GB & Jetson TX2 & Jetson AGX XAvier \\ \hline
\begin{tabular}[c]{@{}l@{}}Host\\ (CPU)\end{tabular} & \begin{tabular}[c]{@{}l@{}}Intel Xeon Silver 4215\\ 8-core 3.5 GHz\end{tabular} & \begin{tabular}[c]{@{}l@{}}ARM Cortex A57\\ 4-core 1.5 GHz\end{tabular} & \begin{tabular}[c]{@{}l@{}}ARM Cortex A57 4-core 2 GHz\\ + Nvidia Denver 2 dual-core 2 GHz\end{tabular} & \begin{tabular}[c]{@{}l@{}}Nvidia Carmel\\ 8-core 2.3 GHz\end{tabular} \\ \hline
\begin{tabular}[c]{@{}l@{}}Device\\ (GPU/DPU)\end{tabular} & DPU 350 MHZ & \begin{tabular}[c]{@{}l@{}}Nvidia Maxwell GM20B\\ 921 MHz\end{tabular} & \begin{tabular}[c]{@{}l@{}}Nvidia Pascal GP10B\\ 1.3 GHz\end{tabular} & \begin{tabular}[c]{@{}l@{}}Nvidia Volta GV10B\\ 1.4 GHz\end{tabular} \\ \hline
Device cores & 2540 cores & 128 Cores & 256 Cores & 512 Cores \\ \hline
Device memory & \begin{tabular}[c]{@{}l@{}}160 GB DDR4\\ 2400 MHz\end{tabular} & 2 GB Shared with system & 8 GB Shared with system & 32 GB Shared with system \\ \hline
Storage memory & \begin{tabular}[c]{@{}l@{}}64 GB DDR4 2400 MHz\\ + 2 TB SSD\end{tabular} & \begin{tabular}[c]{@{}l@{}}2 GB 64-bit LPDDR4\\ 1600 MHz\end{tabular} & \begin{tabular}[c]{@{}l@{}}8 GB 128-bit LPDDR4\\ 1866 MHz\end{tabular} & \begin{tabular}[c]{@{}l@{}}32 GB 256-bit LPDDR4X\\ 2133 MHz\end{tabular} \\ \hline
Power Modes & 23.22 W per \ac{dimm} & 10 W/ 5 W & 15 W/ 7.5 W & 30 W/ 10 W \\ \hline
\end{tabular}}
  \end{center}
  \end{table*}

The experimental values achieved represent the average of $20$ runs. The \ac{gpu} values were obtained by \texttt{clock\_time()} using the MAX-N power mode. In the UPMEM system, the execution time was obtained through an internal cycle counter using \texttt{perfcounter\_config(COUNT\_CYCLES, true)}, and the data transfer times between host and device were obtained using the UPMEM's \ac{dpu} profiling tool, as recommended by the SDK manual~\cite{UPMEM_manual}.

These decoders do not assume early termination and the presented values exclude the processing of the parity check equation verifications (a posteriori information computation and hard decision processing blocks) in order to fix the number of decoding iterations. This design choice ensures a fair and objective comparison of throughput and decoding performance with other works in the literature, independently of channel conditions. Using early termination could bias the results by coupling decoder latency and throughput to the noise level, leading to less consistent performance metrics across different \ac{snr} regimes. While early stopping can reduce average decoding time in practice, its omission here enables controlled benchmarking and avoids introducing conditional branches and thread synchronization overhead on the UPMEM architecture.

\begin{figure}[t]
  \centering
  \includegraphics[width=0.9\columnwidth]{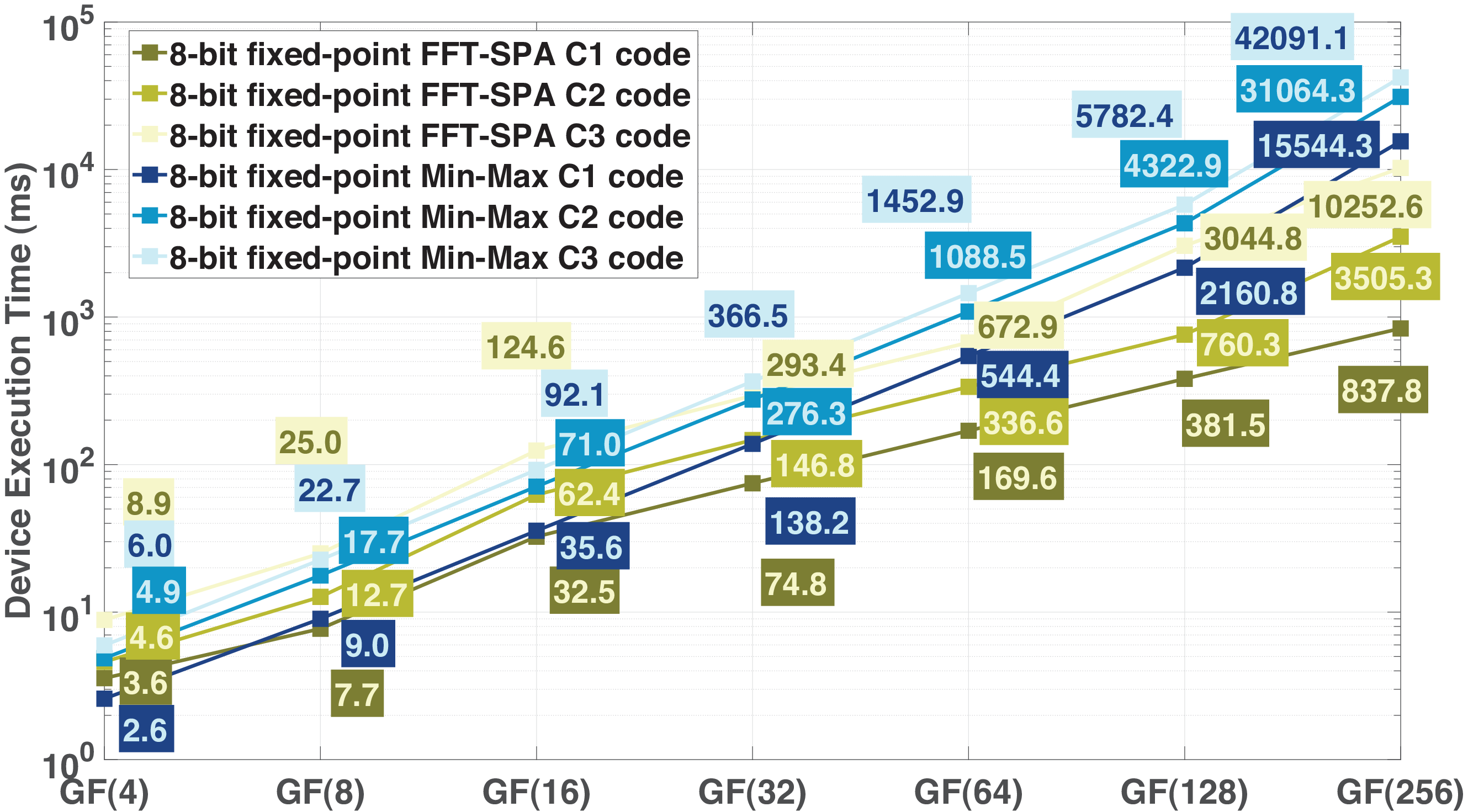}
  \caption{Execution time of \ac{nbldpc} decoding algorithms in a single-threaded \ac{dpu}. The vertical axis is represented in the logarithmic scale. These curves share similarities with Fig~\ref{fig:GF_ops}, where the \ac{fft}-\ac{spa} scales better for higher \acp{gf} than the \ac{mm} decoder.}
  \label{fig:algo_scaling}
\end{figure}

\subsection{Decoding Complexity Validation}

By choosing the \ac{ldpc} code, the theoretical computational complexity of the decoder can be derived for different fields. For the codes used in this work, the expressions of Table~\ref{table:ops} can be computed and are depicted in Fig.~\ref{fig:GF_ops} for the different decoding algorithms. Memory transactions are defined by data movement between memory and registers. 



The \ac{fft}-\ac{spa} has a complexity of $\mathcal{O} \left(Mg\log_2g\right)$ and it scales better for higher \acp{gf}, requiring an order of magnitude fewer operations (both arithmetic operations and memory transactions) for $GF(2^8)$ than the \ac{mm} algorithm, which has a complexity of $\mathcal{O} \left(Mg^2\right)$. However, for $GF(2^2)$ and $GF(2^3)$, the \ac{fft}-\ac{spa} requires more operations than the \ac{mm} algorithm, as shown in Fig.~\ref{fig:GF_ops}. The dashed lines in Fig.~\ref{fig:GF_ops} are higher than the theoretical values due to the arithmetic required for indexing and loop control. 


The analysis of both charts assumes that every type of operation takes the same number of cycles to perform, and every memory transaction has the same latency. In reality, using different types of memory and operations can produce different results. Usually, for \ac{nbldpc} codes, the \ac{mm} decoding can yield better results compared to the \ac{fft}-\ac{spa} since it does not use multiplication and division operations.


\begin{figure}[t]
  \centering
  \includegraphics[width=0.9\columnwidth]{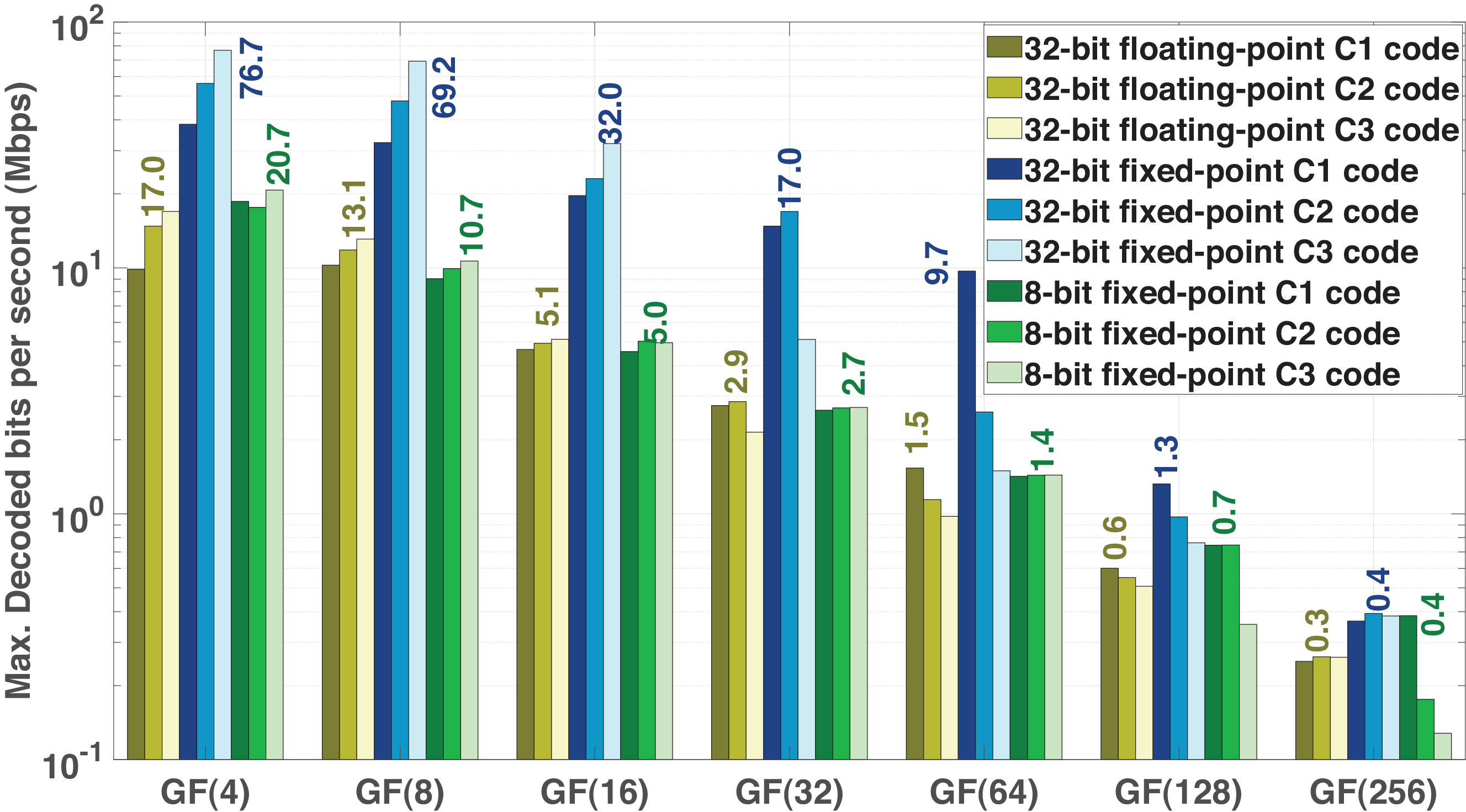}
  \caption{Throughput performance for the multi-\ac{dpu} multithreaded \ac{nbldpc} \ac{fft}-\ac{spa} decoder using $32$-bit integer, and $8$-bit integer representations for three different code sizes. The decoder uses $16$ threads in $2540$ \acp{dpu}. 
  }
  \label{fig:spa_mt_md}
\end{figure}

\subsection{Throughput Performance}

The experimental result shows similarities with the analysis performed in Fig.~\ref{fig:GF_ops}. As depicted in Fig.~\ref{fig:algo_scaling}, the measured execution time in a single-threaded \ac{dpu} shows better scaling for the \ac{fft}-\ac{spa}. The \ac{mm} decoder has a faster execution time for smaller \acp{gf} but it is increases more than the \ac{fft}-\ac{spa} for larger codes and higher \acp{gf}. 

The throughput performance is calculated using $TP=\frac{N\cdot q}{t}$, where $TP$ is the throughput performance measured in decoded bits per second, and $t$ is time. This formula ensures that the number of bits represented in each symbol is taken into account, giving the performance per decoded bit.


\begin{figure}[t]
  \centering
  \includegraphics[width=0.9\columnwidth]{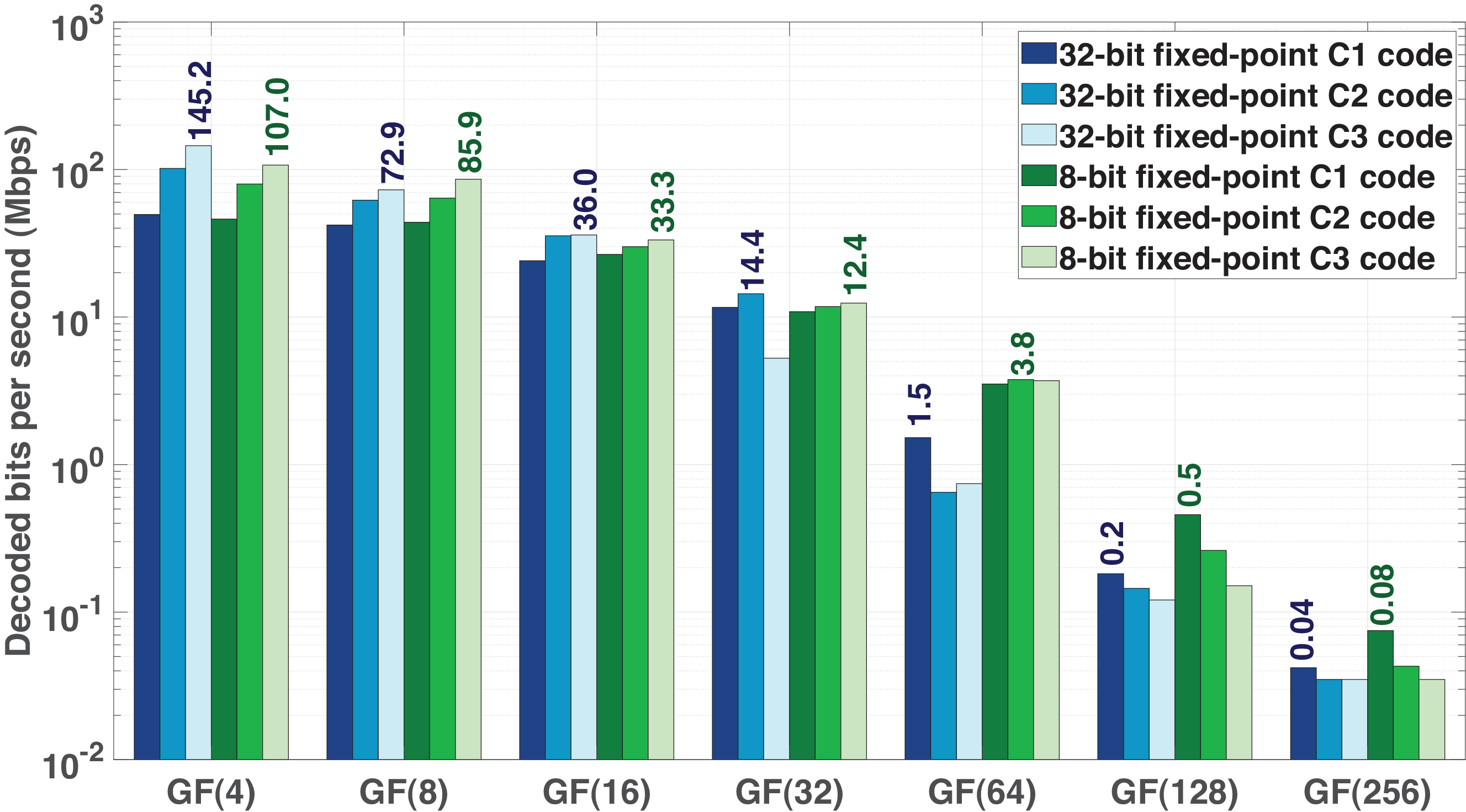}
  \caption{Throughput performance for the multi-\ac{dpu} multithreaded \ac{nbldpc} \ac{mm} decoder using $32$-bit integer and $8$-bit integer representations for three different code sizes. The decoder uses $16$ threads in $2540$ \acp{dpu}. 
  }
  \label{fig:mm_mt_md}
\end{figure}

The following analysis follows a three-step approach to improve performance. The first step uses a single thread to implement the decoder in a single-\ac{dpu}. To improve results, the second step iterates over the first by employing multithreading with $16$ threads. The last step uses multicodewording, where several decoders from the previous step are launched concurrently in $2540$ \acp{dpu} with different codewords to decode.




The throughput values for the \ac{fft}-\ac{spa} with multithreading for $GF(4)$ are presented in Kbps. For $32$-bit \ac{fp}, they are $6.6$, $7.3$, and $7.6$ for $C1$, $C2$, and $C3$ codes respectively. For $32$-bit fixed-point, they are $27.4$, $39.7$, and $54.6$ for $C1$, $C2$, and $C3$ codes respectively. In $8$-bit fixed-point, the values are $28.6$, $45.1$, and $63.6$. The performance decreases from $2.5\times$ to $1.5\times$ for every \ac{gf} increment.  The results show that for small \acp{gf}, the throughput increases for larger codes. However, for higher \acp{gf}, the performance decreases when the code size increases. This performance inversion happens at $GF(32)$ for $32$-bit implementations and at $GF(128)$ for $8$-bit implementations, suggesting that memory usage might impact multithreading implementations. 


The maximum throughput for the $2540$-\ac{dpu} \ac{fft}-\ac{spa} decoders is $16.951$ Mbps using \ac{fp} representation, $76.701$ Mbps using $32$-bit integers, and $20.729$ Mbps using $8$-bit integers, as shown in Fig.~\ref{fig:spa_mt_md}. The measured values represent the execution time between the first \ac{dpu} and the last. 

However, changing from a $32$-bit \ac{fp} to a $32$-bit integer benefits performance. In the single-\ac{dpu} multithreading setup, performance improves between $7\times$ for smaller \acp{gf} and $1.5\times$ for higher \acp{gf}. When comparing multi-\ac{dpu},  performance improves between $4.5\times$ for smaller \acp{gf} and $1.5\times$ .

When comparing the $32$-bit with the $8$-bit implementation for \ac{fft}-\ac{spa}, the performance  in the single-\ac{dpu} multithreading setup improves up to $10\times$. In the multi-\ac{dpu} setup, the performance decreases down to $0.15\times$.




The multithreaded implementation of the \ac{mm} decoder processes different \acp{gf} in parallel. However, the pipeline is underutilized for $GF(4)$ and $GF(8)$, resulting in reduced performance. The throughput values for the \ac{mm} with multithreading for $GF(4)$ are presented in Kbps and, for $32$-bit are $11.8$, $12.6$, and $13.1$ for $C1$, $C2$, and $C3$ codes respectively. For $8$-bit, the values are $11.9$, $12.7$, and $13.1$. For $GF(8)$, the values increase by $25\%$ in average, and for $GF(16)$ the performance decreases by $1.3\times$ compared to $GF(8)$. For higher fields, the performance decreases between $3\times$ and $3.5\times$ on average with each \ac{gf} increment.

The maximum throughput for the $2540$-\ac{dpu} \ac{mm} decoder is $145.2$ Mbps using $32$-bit integer representation, and $107$ Mbps using $8$-bit integers, both in $GF(8)$, as depicted in Fig.~\ref{fig:mm_mt_md}.

For this algorithm, changing from a $32$-bit to an $8$-bit implementation does not provide drastic differences in performance. The single-\ac{dpu} multithreading performance decreases between $1$ and $0.81\times$ for \acp{gf} below $GF(32)$ and improves up to $3.8\times$ for higher \acp{gf}. In the multi-\ac{dpu} implementation, the performance improves and degrades as in the previous observation.

The results for the $8$-bit integer implementations show that \acp{dpu} have underlying issues that prevent the exploitation of the capabilities of the computing units, namely pipeline masking issues since most of the performance is lost when using multithreading for kernels that execute in less than a couple of hundred milliseconds. Higher than this, $8$-bit implementations tend to perform and scale better than their $32$-bit integer counterparts. Furthermore, this effect seems particularly nefarious in multi-\ac{dpu} setups, especially for the \ac{fft}-\ac{spa}, where most kernels execute under $400$ ms.

\subsection{Memory Transactions}

To minimize the impact of costly data transfers between host and device, the chosen approach only transfers the codewords to and from the \acp{dpu} and saves the graph topology in \acp{dpu} without data transfers. The data is always saved in the faster \ac{wram}. However, the limited size of the \ac{wram} does not allow the saving of all data for larger codes in higher \acp{gf}. In this case, the buffer with  fewer memory accesses is saved in \ac{mram} to maximize the number of memory accesses in \ac{wram}.

\begin{figure}[t]
  \centering
  \includegraphics[width=0.9\columnwidth]{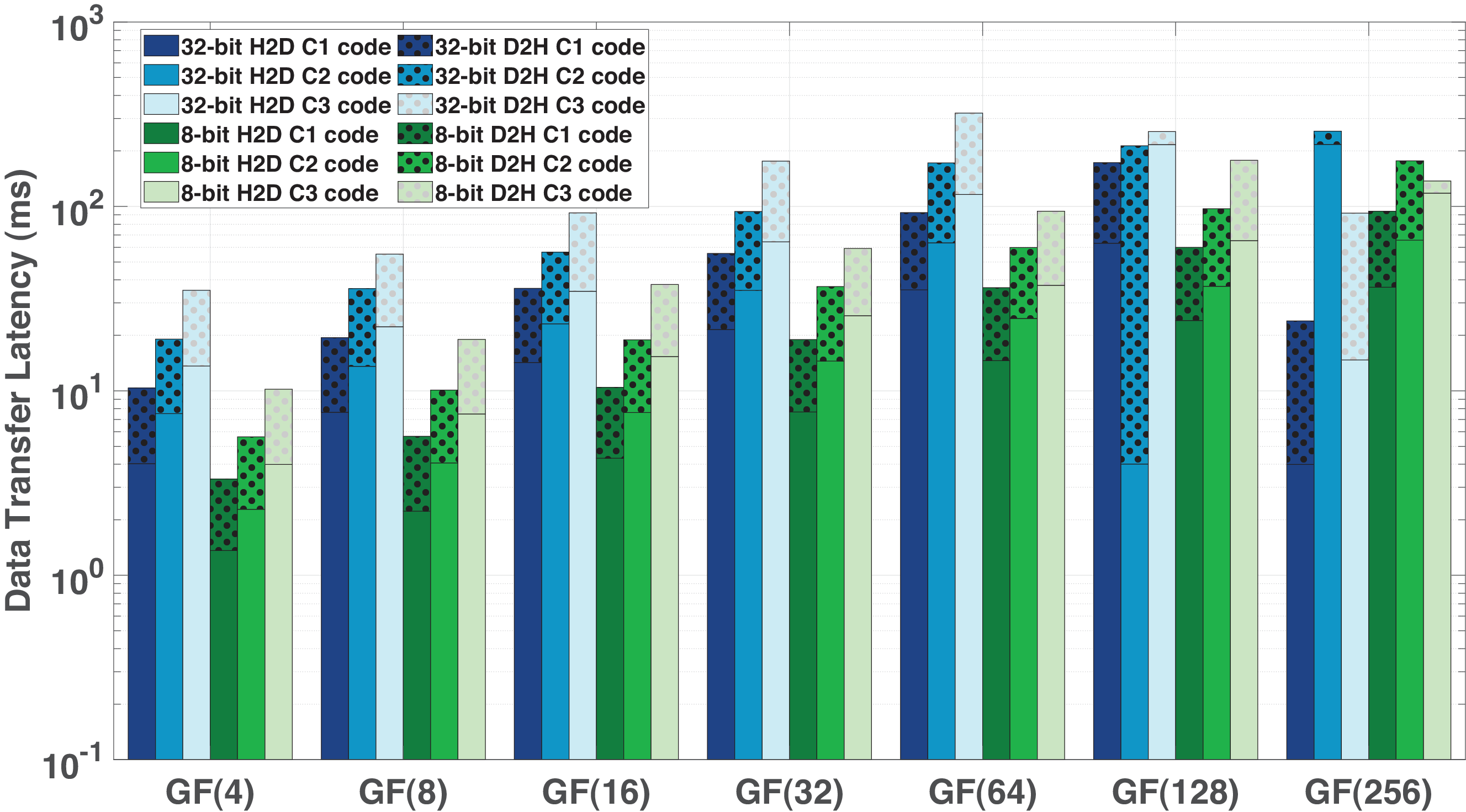}
  \caption{Codeword data transaction time between host and device for the \ac{nbldpc} \ac{fft}-\ac{spa} decoders using $32$-bit and $8$-bit integer codeword representations for three different code sizes. The values represent the transfer times for $2540$ \acp{dpu}. The solid bars represent $2540$ codeword transfers from the \ac{cpu} to the \acp{dpu}. The bars with dots represent the transfer time of all decoded codewords. The values for \ac{mm} decoder are similar since the data size is the same. 
  }
  \label{fig:mem_spa_md}
\end{figure}


The use of \ac{mram} to save some of the buffers implies variations in the execution time for high \acp{gf}. The most prominent cases are shown in Fig.~\ref{fig:mem_spa_md}, for the $32$-bit codeword in the $C2$ code for $GF(128)$ and $GF(256)$ and also in the $8$-bit codeword for $GF(256)$ in $C3$. 


The transfer times of the received and decoded codewords are similar, since the size is the same ($N\cdot 2^q$). For the $8$-bit implementation, the transfer time is reduced $4\times$ for the same code size and \ac{gf}. However, these times are greatly reduced when one or both of these buffers are saved in \ac{mram}. Since \ac{mram} is "closer" to the host, saving data in \ac{mram} is faster than saving data in \ac{wram} which requires the data to travel deeper in the memory hierarchy, taking more time.

For the \ac{fft}-\ac{spa}, Fig.~\ref{fig:mem_spa_md} shows the time when transferring $2540$ codewords for the available \acp{dpu}. For the $32$-bit implementation, in $GF(4)$ for the $C1$ code, it takes between $236$ and $338$ $\mu s$ to transfer one codeword from the host to the device,  and $3.877$ and $6.333$ $ms$ to transfer $2540$ codewords, which means that several codewords are transferred at the same time. The transfer time increases between $10\times$ to $30\times$ when comparing a $32$-bit codeword in single-\ac{dpu} to a multi-\ac{dpu} setup. For $8$-bit codeword, the transfer times increase between $10\times$ to $20\times$. Transferring from device to host also takes the same amount of time. Increasing the code size or \ac{gf} doubles the transfer time (when compared to adjacent \acp{gf} or code sizes).

In Fig.~\ref{fig:mem_spa_md}, the transfer times (for $32$-bit codewords) from host to device (solid bars) for the $C1$ and $C3$ codes in $GF(256)$ are much lower since the buffers are saved in \ac{mram}. The same happens for the $C2$ code in $GF(128)$. For transfers from device to host, $C1$ and $C2$ codes in $GF(256)$ save the output buffer in \ac{mram}, reducing transfer time by $27\times$. For the $C3$ code, the output buffer is saved in \ac{mram} for $GF(128)$ and $GF(256)$.

In the $8$-bit representation, besides reducing the transfer time by $4\times$ compared to $32$-bit, only the output buffer of the $C3$ code in $GF(256)$ is saved in \ac{mram}.

The values measured in the \ac{mm} decoder are similar to those in the \ac{fft}-\ac{spa} where saving data in \ac{mram} also produces the same effect.




\subsection{Comparison to GPUs}

\begin{figure}[t]
  \centering
  \includegraphics[width=0.9\columnwidth]{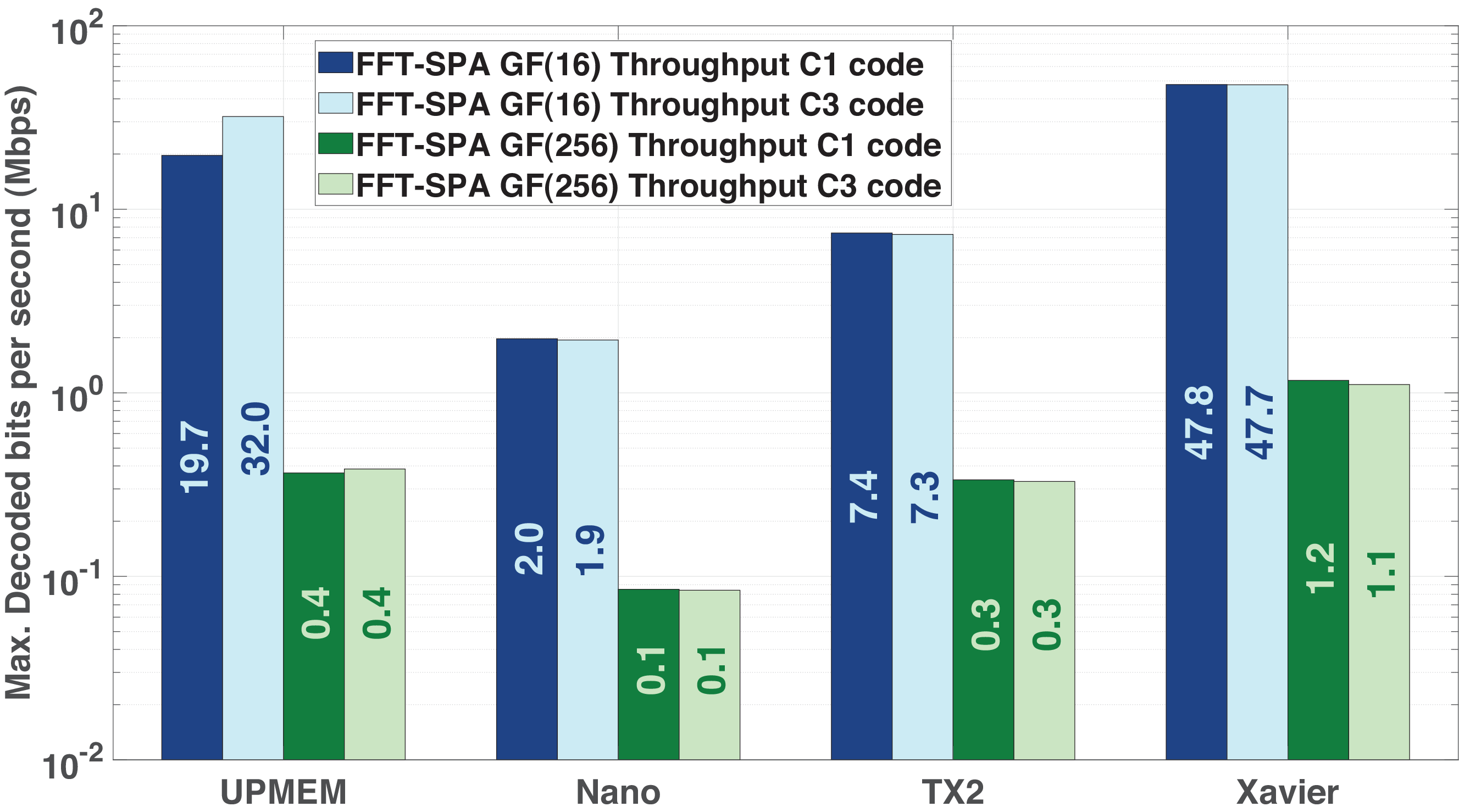}
  \caption{Maximum throughput performance comparison between the UPMEM and low-power \acp{gpu} from~\cite{ferraz:2021:asilomar} using the non-binary $32$-bit fixed-point \ac{fft}-\ac{spa} decoder. The blue bars represent the values for $GF(16)$, while the green bars depict the values using $GF(256)$. 
  }
  \label{fig:comparison_nb}
\end{figure}

Fig.~\ref{fig:comparison_nb} compares the \ac{fft}-\ac{spa} between UPMEM and low-power \acp{gpu} using a multithreading and multiple \acp{dpu} $32$-bit integer implementation. The UPMEM implementation can achieve $32.043$ Mbps for $GF(16)$ and $385$ Kbps for $GF(256)$. The \ac{gpu} values from~\cite{ferraz:2021:asilomar}  also present the maximum extractable performance, running $16384$ decoders in parallel for the Jetson Nano, and $32768$ decoders for TX$2$ and Xavier.

From this figure, the UPMEM system can be compared to a low-power  \ac{gpu} somewhere between a TX$2$ and Xavier. However, these \ac{gpu} implementations can run more decoders in parallel, while the performance per core in the UPMEM system is higher than in \acp{gpu}.

The limited nature of the UPMEM system, and \ac{pim} systems in general, does not allow the exploitation  of some techniques used in \acp{gpu}. For instance, ordering the data has more benefits in \ac{gpu}-based implementations, allowing coalesced memory accesses where several threads in the same warp  access memory simultaneously. If memory accesses are not contiguous, threads in the same warp cannot access memory simultaneously, increasing overall memory access time. 

In \ac{ldpc} codes, coalescing can be leveraged to increase performance by ordering the messages transmitted  between the nodes. However, ordering the data for reading and/or writing for one type of node implies that the reads/writes for the other type of node are not contiguous. Coalescing mechanisms that allow the exploitation of bulk data transfers, are not present in the UPMEM system.

Furthermore, the UPMEM system does not feature caching mechanisms, unlike \acp{gpu}, that allow faster data access to increase performance. Instead, UPMEM features the \ac{wram} which is controlled by the programmer resulting in average faster slower data access times. 

\section{Key Takeaways}

These decoder implementations and experimental results of this research, highlight the implications for in-memory \ac{ldpc} decoding and potential areas for future exploration for both parallel computing and coding theory communities. The main findings and observations of this work are: 

\begin{itemize}

\item \textbf{First \ac{pim}-Based \ac{nbldpc} Decoder Implementations}: The paper introduces the first known \ac{pim}-based implementations of the \ac{fft}-\ac{spa} and the \ac{mm} algorithm for \ac{nbldpc} decoding. 

\item \textbf{\protect\ac{pim} Technology for \ac{ldpc} Decoding}: The use of \ac{pim} technology, specifically the UPMEM system, can significantly alleviate the data movement bottleneck in \ac{nbldpc} decoding, enhancing overall performance. 

\item \textbf{Competitive Throughput}: The \ac{pim}-based \ac{nbldpc} decoder achieves a decoding throughput of $76$ Mbit/s, making it competitive with edge \ac{gpu} implementations. 

\item \textbf{Optimization Techniques}: Effective optimization techniques, such as maximizing \ac{wram} usage, quantization, and loop unrolling, are crucial for enhancing the performance of \ac{ldpc} decoders on the UPMEM system.

\item \textbf{Scalability and Modularity}: The modular nature of the UPMEM system allows for scalable performance improvements by adding more \ac{pim} modules, unlike \acp{gpu} which require complete device replacements.

\end{itemize}


Efficient parallel computing requires understanding both software and hardware to fully exploit device capabilities. This work applies \ac{pim} technology to \ac{ldpc} decoding and offers recommendations for the parallel computing community:

\begin{tcolorbox}[colback=blue!5!white,colframe=blue!75!black,fonttitle=\small,title=KEY TAKEAWAYS FOR PARALLEL PROGRAMMERS]

1. \textbf{\protect\ac{wram} Usage}: Prefer using \ac{wram} over \ac{mram} for data storage, except when data has low reuse and needs to be transferred to the host.  \ac{wram} offers faster access times and higher bandwidth compared to \ac{mram}. 

2. \textbf{Quantization}: Use quantization schemes and integer or fixed-point arithmetic instead of \ac{fp} arithmetic to reduce complexity, power consumption, and memory footprint, while maintaining acceptable error-correction performance. 

3. \textbf{Loop Unrolling}: Manually unroll loops to remove branching operations from the \ac{alu}'s pipeline, reducing the total number of operations and improving performance at the cost of a higher memory footprint. 

4. \textbf{Multithreading}: Utilize multithreading to fully exploit the \ac{dpu} pipeline by distributing the load over more than $11$ threads, ensuring efficient parallel processing. 

5. \textbf{Memory Management}: Explicitly manage data transfers between \ac{mram} and \ac{wram} using \ac{dma} instructions to minimize costly data transfers and maximize the use of faster \ac{wram} for frequently accessed data. 

\end{tcolorbox}

\begin{tcolorbox}[colback=green!5!white,colframe=green!75!black,fonttitle=\small,title=KEY TAKEAWAYS FOR HARDWARE DEVELOPERS]

1. \textbf{Reducing Complexity}: By converting \ac{fp} arithmetic to integer or fixed-point arithmetic, quantization simplifies the arithmetic operations, making them less complex and faster to execute. 

 2. \textbf{Enhancing Speed}: Integer and fixed-point operations are generally faster than \ac{fp} operations, especially on hardware like the UPMEM system, which lacks native floating-point units and emulates these operations in software.


3. \textbf{Minimizing Memory Footprint}: Quantized data requires less memory compared to \ac{fp} data, allowing more efficient use of available memory resources. While there is a small decrease in error-correction capability, the trade-off is often acceptable given the significant gains in speed, power efficiency, and reduced complexity. 

\end{tcolorbox}

\section{Conclusion}

This work highlights the potential of general-purpose \ac{pim}-based acceleration for \ac{nbldpc} decoding using the UPMEM system. While individual \ac{dpu} cores in the UPMEM system are limited in performance, the parallel architecture and proximity to memory provide a significant advantage by reducing data movement bottlenecks. Although the UPMEM system may not fully match edge \acp{gpu}, it reaches performance levels comparable to low-power \acp{gpu} when executing fewer parallel decoders, indicating that performance is primarily constrained by the number of computing cores.

Furthermore, the modularity of the UPMEM system allows scalable performance improvements simply by adding more \ac{pim} modules, unlike \acp{gpu} which require complete device replacements for upgrades.

Overall, message quantization helps in achieving a balance between performance and resource utilization in \ac{ldpc} decoders, in particular those exploiting \ac{pim}-based functional units. Furthermore, this work demonstrates the feasibility of \ac{pim}-based \ac{nbldpc} multicodeword decoding and provides valuable insights for future research, including new capability of processing larger datasets while moving less data, the adoption of more aggressive process node designs for supporting more complex operations, opening the door to broader applications like machine learning or image processing in this novel computing domain.

The proposed architecture is particularly well suited to high-density batch decoding environments, such as massive \ac{mimo} base stations, optical communication receivers, and cloud-\ac{ran} infrastructure, where decoding hundreds or thousands of codewords in parallel is essential to meet data rate demands, despite the initial latency.

\section*{Acknowledgments}
This work was supported by Instituto de Telecomunicações
and FCT - Fundação para a Ciência e Tecnologia, I.P. by
project reference 10.54499/UIDB/50008/2020, and DOI identifier
https://doi.org/10.54499/UIDB/50008/2020, UIDP/50008/2020,
2022.06780.PTDC and Ph.D. grant 2020.07124.BD.

\bibliographystyle{unsrt}  
\bibliography{references}

\end{document}